# A Critical Review of Baseband Architectures for CubeSats' Communication Systems

Amr Zeedan and Tamer Khattab, Senior *Member, IEEE*

***Abstract*— Small satellite communications recently entered a period of massive interest driven by the uprising commercial and civil space applications and motivated by various technological advances. Miniaturized satellites, known as CubeSats, are particularly attractive due to their low development and deployment costs which makes them very promising in playing a central role in the global wireless communication sector with numerous applications ranging from Earth imaging and space exploration to military applications. Moreover, constellations of CubeSats in low-earth orbits can meet the increasing demands of global-coverage flexible low-cost high-speed connectivity. However, this requires innovative solutions to overcome the significant challenges that face high-data-rate low-power space communications. This paper provides a comprehensive and critical review of the design and architecture of recent CubeSat communication systems with a particular focus on their baseband architectures. The literature is surveyed in detail to identify all baseband design, testing, and demonstration stages as well as accurately describe the systems' architecture and communication protocols. The reliability, performance, data rate, and power consumption of the reviewed systems are critically evaluated to understand the limitations of current CubeSat systems and identify directions of future developments. It is concluded that CubeSat communication systems still face many challenges, namely the development of energy-efficient high-speed modems that satisfy CubeSats' cost, mass, size, and power requirements. Nevertheless, there are several promising directions for improvements such as the use of improved coding algorithms, use of Field Programmable Gate Arrays, employment of multiple access techniques, employment of beamforming and Multiple-Input Multiple-Output techniques, use of advanced antennas, and transition to higher frequency bands. By providing a concrete summary of current CubeSat communication systems designs and by critically evaluating their unique features and limitations as well as offering insights about potential improvements, the review should aid CubeSat developers, researchers, and companies working in the field to develop more efficient and high data rate CubeSat systems.**

## I. Introduction

Satellite communications play a critical role in the global telecommunications systems especially with the development of smart cities and the increasing use of Internet of Things (IoT) platforms. There are approximately 2000 satellites, excluding nanosatellites and CubeSats, orbiting the earth that exchange wireless signals carrying various types of data (voice, image, etc.) to and from many locations worldwide [1]. Conventionally, communication satellites are large and have geosynchronous orbits. However, recently, many companies and organizations are moving towards building constellations of smaller and cheaper satellites (e.g., CubeSats) in low-earth orbits (LEOs) that can provide communication services to both urban and rural areas and meet the increasing demand of low cost and high data rate connectivity. CubeSats promise more flexible services, shorter development times, and lower development and operation costs. However, providing high speed communication services using CubeSats requires innovative solutions to overcome the significant hurdles that face this idea; mainly power consumption, data rate, and form factor (size and weight) constraints. These constraints require solutions that are largely contradicting and very challenging to be all simultaneously achieved. Consequently, there have been many active research efforts in developing and optimizing CubeSat communication systems especially in recent years where achieving high data rates is becoming increasingly important to meet the demands of the commercial applications of CubeSats.

Although there are several review papers on CubeSats and satellites' communication systems [2]– [7], there are little or no comprehensive reviews on the design details or architectures of CubeSats' communication systems. So, this paper will fill in this gap in the literature by reviewing and evaluating recent CubeSat communication systems, with a particular focus on their baseband architectures, and their resulting performance in terms of power consumption, data rate, and error rate. The paper aims to provide a well-grounded starting point for researchers and developers working on custom designs for CubeSat communication systems.

This publication was made possible by an award GSRA8-L-2-0523-21052 from Qatar National Research Fund, QNRF (a member of Qatar Foundation, QF), supporting A. Zeedan, and award number AICC03-0530-200033 from QNRF (a member of QF), supporting T. Khattab. The contents herein are solely the responsibility of the authors. *(Corresponding author: Amr Zeedan).*

Amr Zeedan and Tamer Khattab are with the Department of Electrical Engineering, Qatar University, Doha, Qatar (e-mail: az1706240@qu.edu.qa; tkhattab@ieee.org).

TABLE I
COMPARISON BETWEEN THIS REVIEW PAPER WITH PREVIOUS REVIEWS.

| Reference | Year | Scope |
|---|---|---|
| Rahmat-Samii et al. [2] | 2017 | Antenna designs and developments for CubeSats. |
| Gao et al. [3] | 2018 | Developments in advanced antennas for small satellites. |
| Davoli et al. [4] | 2018 | CubeSat mission goals, constellation topologies, and communication protocols. |
| Burleigh et al. [5] | 2019 | Operational features of small satellites, CubeSat services and applications, and network protocols. |
| Saeed et al. [6] | 2020 | CubeSat constellations and coverage, channel modelling, link budget, and networking and upper layer issues. |
| Kodheli et al. [7] | 2021 | Advances in satellite (mainly large satellites) communications, applications, medium access, networking, and prototyping. |
| This paper | 2022 | Baseband designs, architectures, and technologies of CubeSat communication systems, power consumption and data rate investigation, performance and design evaluation, limitations of current systems, and directions for future developments. |

This will be achieved by reviewing current systems, their designs, and architectures as well as by comparing their performance and pointing out their limitations and potential future improvements. This will lead to a firm understanding of the current state of the art technologies in the field and pave the path for introducing improvements and upgrades on current systems.

A significant challenge in conducting this review is the fact that most papers in the field focus more on the RF-end design, link budget, and bandwidth requirements than on the design of the baseband system and its optimization and power requirements. Consequently, a very detailed investigation of the existing literature is needed to extract the baseband architecture designs and evaluate them. Table I provides a comparison between this review paper and several previous reviews on CubeSats, illustrating the uniqueness of this review. As it can be noted, the contribution of this review is three-fold. First, the paper provides a comprehensive and detailed review of recent CubeSat communication system designs focusing on the baseband architectures and corresponding algorithms that optimize the performance of the baseband stage. The different systems are categorized into four categories, as will be explained later, based on the adopted design approach. Second, the paper provides critical evaluation of the reviewed systems to accurately compare their performance and point out to their limitations or shortcomings. Third, the paper provides insights into future trends, technologies, and directions for CubeSat communications development both from a research perspective and commercial applications perspective.

This paper is organized as follows: a compact background is given in section 2 to provide the reader with the necessary background knowledge about nanosatellites, their development and applications, the general architecture of digital communication systems, the main challenges that face nanosatellites communications, and a brief overview of FPGAs (Field Programmable Gate Arrays) and SDRs (Software Defined Radios) as many of the works reviewed here employ them in their designs. Section 3 constitutes the primary literature review of the baseband architectures used for CubeSats' communication systems. In the fourth section, the performance of the reviewed systems will be evaluated and compared in terms of power consumption, data rate, frequency bands, and other elements. Section 5 is a critical discussion of the current systems, their limitations, and potential directions for future developments. The last section summarizes the main lessons learned from this review and concludes the paper.

## II. BACKGROUND

### *A. Development and Applications of Nanosatellites*

Nanosatellites are defined as satellites with masses between 1 kg to 10 kg [8], [9]. The term "CubeSat" refers to the size and shape of the nanosatellite being a cube with dimensions of 10 cm × 10 cm × 10 cm or called 1U [10]. CubeSats can be 1U, 2U, 3U, or 6U in size and should have a mass up to 1.33 kg per unit [10], [11]. The terms nanosatellites and CubeSats are used here interchangeably despite the technical size and mass definition difference between them. There have been around 1600 nanosatellites launches between 1998 and 2021 [8]. Nanosatellites are used for various applications including Earth observation, remote sensing, technology demonstration, communication services, military applications, commercial applications, and interplanetary space exploration [5]. Fig. 1 shows the number of nanosatellites launched per year by their intended area of application. There has been a tremendous increase in the number of nanosatellites launched in the last few years. It is seen that in the first years of nanosatellites launches, most of the satellites were intended for research purposes and technology demonstration

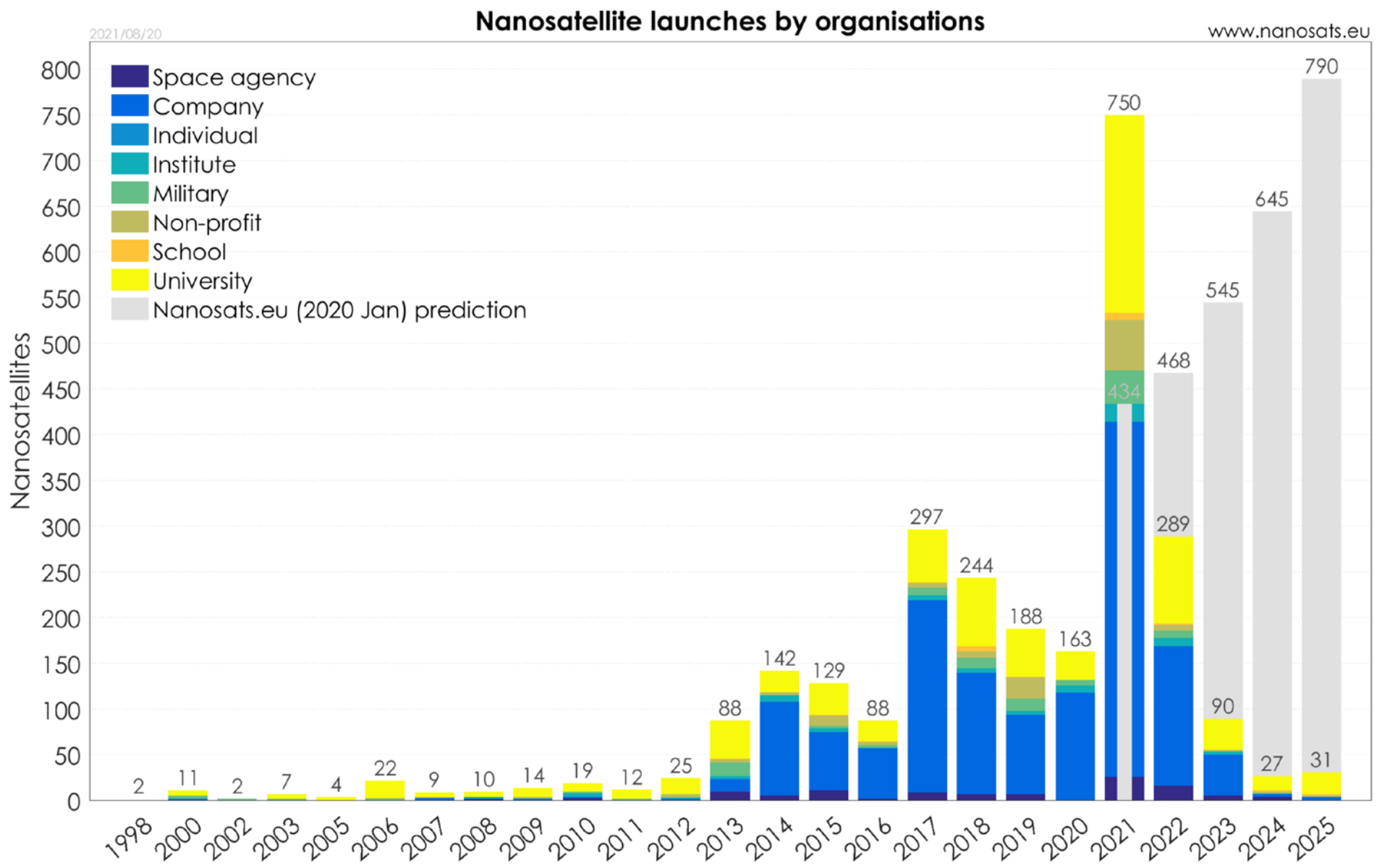

**Fig. 1.** Number of nanosatellites launches per year categorized by use [8].

[12]. However, starting from around 2014 more and more nanosatellites are being launched for commercial purposes. This indicates that nanosatellites are no longer exclusively being used for research and earth observation applications but also for providing telecommunication services, competing with traditional satellites [5]. Consequently, the development of small-size high-speed energy-efficient reliable communication systems for CubeSats is becoming ever more critical to serve the anticipated telecommunication needs. Due to the very small size, mass, and consequently limited energy resources available on nanosatellites, the communication systems used must be highly efficient. Moreover, while for research nanosatellites the data rate may not necessarily need to be high, telecommunication nanosatellites will be required to have very high data rates to meet the increasing demands of users who are looking for high speed connectivity especially with the increasing commercial use of IoT services, smart cities, and other emerging telecommunication technologies. The enterprise of designing such communication systems is thus of critical importance both from a research point of view and technological/ commercial point of view.

One of the main motivations for the increased interest in CubeSats is their low cost compared to large satellites. A typical conventional satellite would cost between 150 to 350 million dollars [4]. A CubeSat on the other hand costs less than 200,000 dollars [4]. In addition to the considerably lower cost, this makes CubeSats more accessible to companies of different types and sizes, which is reflected in the considerable increase in the number of civil and commercial CubeSat operators in recent years [14]. This is aside from the much shorter development times needed for developing and launching a CubeSat which could be under a year, while developing and launching a large satellite requires between 5 to 15 years [4], [13]. However, this is at the cost of shorter lifetimes for CubeSats in space which are between 2 to 4 years of operation, but at the same time this gives the opportunity for incorporating up to date technologies in the CubeSat that replaces the old one in the constellation [13]. Thus, the telecommunication services offered by CubeSats can be significantly improved by constantly making use of latest technologies. CubeSats also feature several other advantages over conventional satellites such as risk distribution, flexible services, smaller sizes, lower communication latency, lower energy consumption, and lower human power needed [4], [13].

### *B. General Architecture of Digital Communication Systems*

A typical wireless digital transmitter consists of a source encoder, channel encoder, digital modulator, and RF antenna [15], [16]. The transmitter can have additional blocks such as a multiplexer, frequency spread, and multiple access. On the receiver side, there is an RF antenna, demodulator, channel decoder, and source decoder [15], [16]. If the original message is analog, then there will also be an analog to digital converter (ADC) before the source encoder and a digital to analog converter (DAC) after the source decoder. Before a given message can be transmitted it must go through different stages of signal processing both at the baseband level (low frequency) and bandpass level (high frequency) [16]. The functional blocks at the transmitter are responsible for processing the original message, source and channel encoding, modulation, and transmission over the wireless channel [16]. The functional blocks at the receiver perform the reverse processes to reconstruct the original message. Fig. 2 illustrates the basic blocks of a digital communication system, which are part of the design of any modem. In the first block, the source message is converted into a suitable data format; digitalized using ADC if it is an analog signal. Then, the message is source encoded to reduce its size without losing any of the original information, ideally [15]. The message could also be subjected to encryption with specific ciphers or keys that are only known to the transmitter and receiver to protect the confidentiality of the message. After that, the channel encoder adds redundant bits to allow for error detection and correction at the receiver after the signal travels through the noisy channel [16]. The signal is then modulated using some digital modulation scheme and shifted to a higher frequency to be radiated through the RF antenna. On the receiver side, the RF signal is captured by the antenna and then converted back to low frequency by demodulation. The channel decoder decodes the message with the help of the redundant bits in order to detect the transmission errors and correct as many of them as possible [16]. Then, decryption and source decoding are performed to reconstruct the original message which is then converted back to its original format (e.g., analog) [16]. There can be additional blocks that process the signal further before transmission or after reception to increase the signal quality or improve the utilization of resources (e.g., power, bandwidth, etc.). The focus of this paper is on the baseband part of the communication system, which includes the blocks before the final RF processing and transmission of the signal.

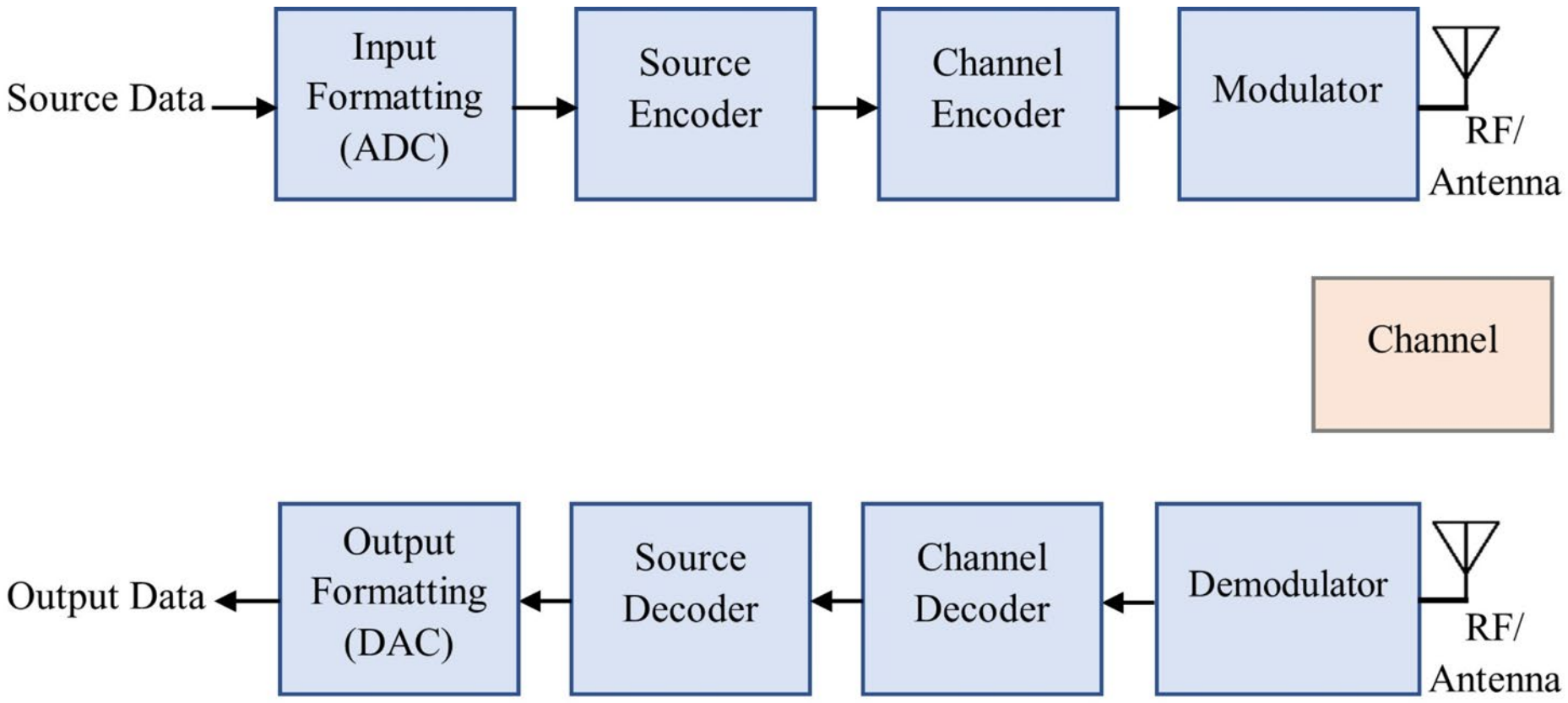

**Fig. 2.** Basic blocks of a digital communication system [15].

### *C. Challenges of Designing CubeSats' Communication Systems*

A high-quality wireless communication system needs to have high data rate, high spectral or bandwidth efficiency, high power efficiency, high signal quality (low probability of error), and low-cost implementation [17]. Designing an efficient (in terms of resources such as energy and bandwidth) and reliable (in terms of data rate and signal quality) wireless communication system is a difficult task in its own right. This is because attempting to optimize one of these features automatically results in the degradation of another [17]. For instance, increasing the data rate conventionally requires increasing the utilized bandwidth, and improving the quality of communication typically requires higher transmitted signal power [17]. Therefore, it is difficult to design a very reliable communication system that is highly resource-efficient at the same time; there is always a tradeoff in wireless communications. This challenge becomes even more elevated for the communication system of a CubeSat due to the additional constraints that this system must satisfy. Firstly, the very small size and mass of the CubeSat limits the physical dimensions of the communication system [3]. That is, every part of the system must be designed to be as compact and as light as possible including the antenna. Secondly, due to the small size of the CubeSat, the available energy resources such as batteries and solar panels are very limited, and hence, the generated power has to be distributed efficiently over the different sub-systems of the CubeSat [3]. Thus, the power requirements of the communication system must be very low. At the same time, this cannot come at the expense of low data rate or poor communication quality, especially if the CubeSat is intended for providing communication services. In such case, the quality and data rate are essential to the success of the service provider. Another constraint is the low-cost requirement of the CubeSat [3]. Since a major feature of CubeSats is their low development costs, the components and materials that go into the implementation of the communication system cannot be very expensive. So, the communication system must be cost-efficient to a large degree. Moreover, the materials used to construct the communication system must be chosen very carefully to survive the harsh space environment [3]. The materials used must be able to withstand the various thermal and radiation effects of the orbital environment [3].

CubeSats' communication systems have developed considerably over the years of CubeSat missions. In the early years of CubeSat, data rates were very low ranging from 1 to 10 kbps [5]. Amateur frequencies were used for communications at the ultrahigh frequency (UHF) band (0.3-1 GHz) of the electromagnetic spectrum [18]. However, as CubeSats evolved from being used primarily for research purposes to other emerging commercial and civil applications, their communication systems witnessed rapid developments in almost all aspects. Higher frequency bands such as the S-band (2-4 GHz) and X-band (8-12 GHz) became more widely used due to the developments in the commercially available monolithic microwave integrated circuits [5]. With the use of such high frequency bands, data rates were able to reach 100 kbps to 1 Mbps [5]. Higher data rates require utilization of higher frequency bands like Ku (12-18 GHz), K (18-27 GHz), and Ka-band (27-40 GHz) [4]. These bands are still emerging technologies for small satellites due to the types of antennas needed and the power requirements [5]. Another major aspect of development is the increasing move towards digital implementation of the communication system or SDR, which is explained in the following sub-section. This transition is driven by the need for reconfigurable and flexible radio communications and by the advances in the corresponding enabling technologies such as digital signal processors (DSPs) and FPGAs [5]. The challenge of this digital approach is the power consumption constraint of small satellites. That is why FPGAs are preferred for implementation, especially for higher data rates in the Ka-band since they can perform computationally intensive tasks in parallel and with better efficiency every clock cycle [19]. It is the aim of this paper to review such state-of-the-art CubeSat communication systems, mainly focusing on the baseband design and architectures, that can provide high data rate links.

### *D. Brief Background on FPGAs and SDR*

In this sub-section, a brief overview of FPGAs and SDR will be given as they form a major part of the design of wireless communication systems for small satellites as will be seen in the review. This sub-section aims to provide sufficient background about these two concepts so that the non-expert reader can follow up with the extensive review that will be presented in the next section.

A Field Programmable Gate Array (FPGA) is an integrated circuit (IC) that can be configured for a specific application by the designer [20]. An FPGA contains thousands of configurable logic blocks (CLBs) including look-up tables, multiplexers, and flip-flops that are surrounded by programmable interconnections which can be reconfigured by the designer to implement different logical functions [20]. In other words, on an FPGA, the designer programs the hardware of the device rather than writing the software on a predefined processor. That is why FPGAs are used primarily to design application-specific integrated circuits (ASICs). FPGAs allow prototype testing and optimization of the proposed circuit architecture before a hardware-fixed ASIC is manufactured [21]. This approach greatly saves time and money since it allows testing many versions of the same project before the final ASIC manufacturing process [20]. FPGAs are programmed by hardware description languages (HDLs) such as VHDL and Verilog [21]. Using an FPGA in designing a wireless communication system architecture is not only useful in optimizing the chip for the particular workload required but also for having reconfigurability while employing the system [21]. For example, transitions between different modulation or coding schemes can be realized when the system is implemented on an FPGA. Thus, the communication system can be made more efficient by switching to different communication architectures based on the needs of the users and by employing the same hardware resources. Moreover, the response time can be controlled on an FPGA if compared to a standard central processing unit (CPU) [21]. The required algorithm that ensures a fixed and short response time can be implemented on the FPGA and then be employed after testing [21]. An IC designed this way will usually have a quick response time and efficient power consumption, which are characteristics critically sought after in the communication systems of CubeSats. Thus, FPGAs provide several important features for designing and implementing efficient wireless communication systems and that is reflected in their wide use in the works that will be reviewed in this paper.

Software Defined Radio (SDR) is an important technology used in wireless communications. SDR is a radio frequency (RF) communication system where RF communication is achieved using software to perform the signal processing tasks that are conventionally performed by hardware [22]. That is, in SDRs, blocks like modulators, detectors, and mixers are implemented by means of software programming on a computer or a reconfigurable digital electronic board rather than on multiple hardware components [22]. For instance, in an SDR transmitter, the software should typically be responsible for generating the baseband, intermediate frequency (IF), and RF waveforms. Similarly, an SDR receiver will have the demodulation and decoding performed using software rather than hardware components. An SDR utilizes a high-speed reconfigurable hardware device, such as an FPGA or a DSP, that can perform many functions based on the software program loaded [23]. A communication system implemented on an FPGA is a practical example of an SDR system that possess all the reconfigurability and efficiency features explained in the previous paragraph. FPGAs' configurability and speed have made them a common platform for realizing SDRs [24]. However, the SDR flexibility comes at the cost of increased software complexity due to the various algorithms that need to be programmed for code generation, debugging, etc. [23], [24]. They are also generally more expensive than single-chip highly integrated transceivers, which provide a completely hardware-based RF system [23]. Therefore, SDRs are not necessarily the best option for designing a wireless communication system, a completely hardware radio or a mixed hardware and software design may offer a more efficient design depending on the specific application of the communication system. This will be evident from the works that will be surveyed in the next section and their corresponding performance evaluation that will be carried out in the fourth section.

## III. BASEBAND ARCHITECTURES IN THE LITERATURE

There are various approaches for designing a CubeSat communication system. One of the most common and trending approaches is to utilize the concept of SDR to custom-design a reconfigurable communication system that meets the specific requirements of the CubeSat mission [25]. Such designs are categorized as custom designed SDRs. In this approach, the designer typically chooses the suitable processor for the system, the transceiver, DAC/ ADC, and other components such as amplifiers. This also involves customizing and developing the architecture of the system and the algorithms used for the different functions. Another design approach is to base the communication system on some already existing SDR platform that is commercially available [25]. So, the required components and resources are already existing, and the design is more concerned with utilizing the available resources to optimally implement the desired application. This is categorized as commercial SDR. The more conventional approach is to use hardware components to implement the communication system. As is the case with SDR, there are also custom hardware design approaches and commercial hardware designs, where in the later the hardware components used are consumer off-the-shelf components rather than custom designed hardware components. The review is divided into four sub-sections based on the design approach category. In each sub-section, the relevant works falling under that category are reviewed in terms of the communication system design and implementation with a particular focus on the baseband architecture. In the next section, the performance of all the reviewed systems will be evaluated and compared. It is worth emphasizing that this review only includes relatively recent designs (starting from around 2016) that have a minimum data rate of 100 kbps. However, few older designs are reviewed for comparison purposes.

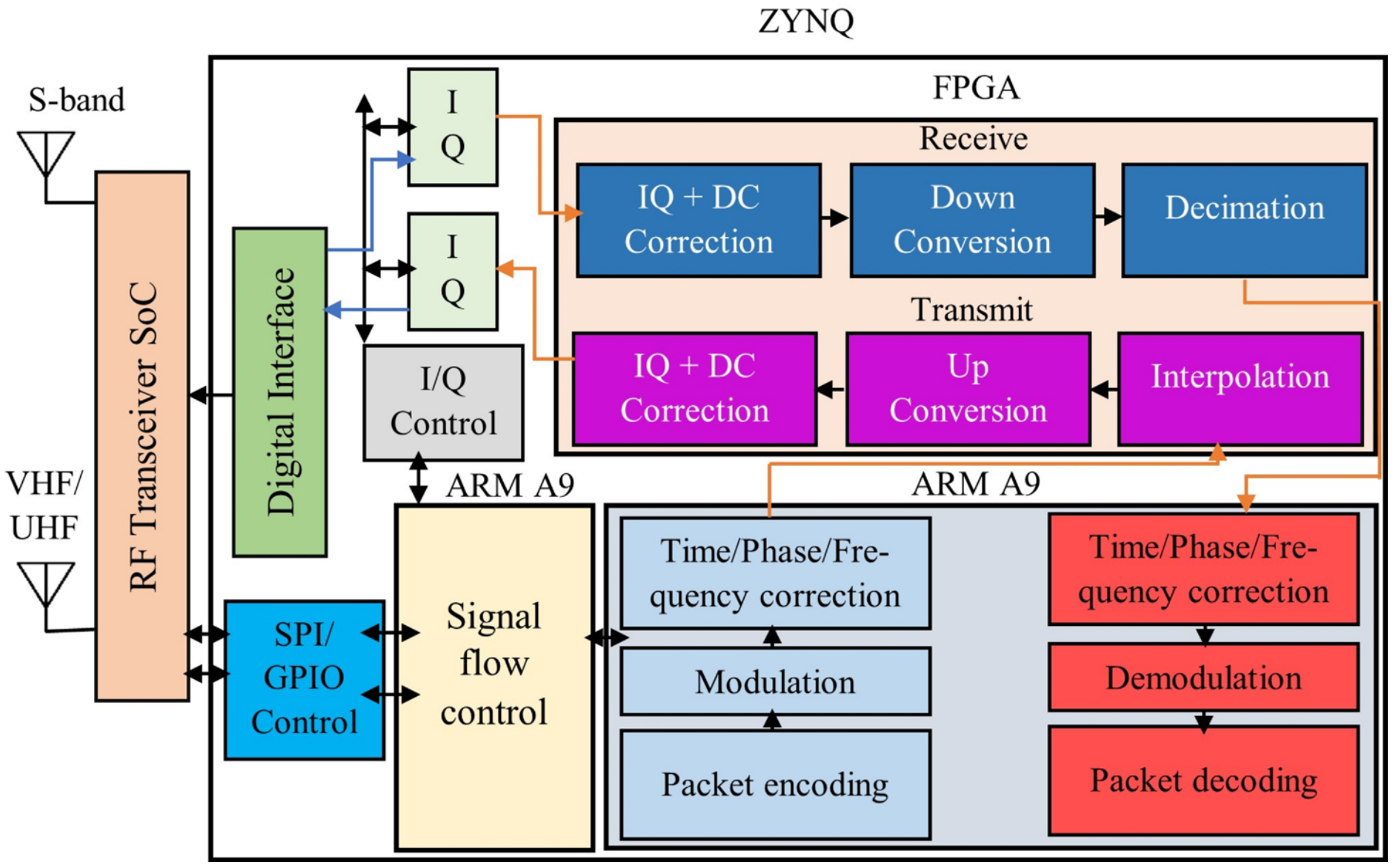

**Fig. 3.** Multi-core SDR architecture used in [27].

*A. Custom Designed SDRs*

FPGAs are widely used in advanced state-of-the-art SDR designs for CubeSats communication systems [19]. A typical implementation of SDRs is to program a DSP or an FPGA to execute all functions of filtering, error correction, framing, etc. [19]. Such functions are computationally intensive even for modern processors to efficiently utilize the available bandwidth [19]. FPGAs are able to execute such intensive tasks in parallel and more efficiently increase throughput while maintaining low power with every clock cycle [19]. Another disadvantage in using DSPs is that as the frequencies used reach S-band and above, they become inefficient in keeping up with executing the necessary functions such as filtering, encryption, and error correction [19]. In such cases, using multi-core processors becomes necessary to achieve the required throughput. However, this adds to the cost, size, power, and complexity of the system. Due to these various problems in using DSPs, Marshall Space Flight Center's SDR system, named Programmable Ultra Lightweight System Adaptable Radio (PULSAR), utilizes an FPGA in its implementation [19]. All the signal processing is performed on the FPGA using HDL. The modulation scheme used in PULSAR is QPSK (Quadrature Phase Shift Keying), operating in the S-band at a data rate of 5-10 Mbps [19]. Due to the flexibility of PULSAR design, it can operate using various types of channel coding schemes based on the requirement of the mission. These schemes are Low Density Parity Check (LDPC), convolutional (rate ½), and Reed-Solomon (255/223) Forward Error correction (FEC) codes [19]. The transmitter has digital algorithms to perform FEC and NEN (Near Earth Network) compatible packetization and the receiver has algorithms for performing signal recovery and error correction [19], [26]. Although this SDR implementation provides flexibility, small size, and low power consumption, it requires very intensive computations at higher data rates and frequencies. The PULSAR SDR consumes about 1 W per each 10 Mbps of data rate [19]. Therefore, for its S-band data rate, it consumes between 0.5- 1 W. Also, PULSAR has an implemented X-band transmitter, but not with an X-band receiver so not a complete system yet, that will transmit one channel of QPSK at a data rate of 110 Mbps but the S-band receiver will only be able to receive that data at 300 kbps [19]. At the current power performance of PULSAR, the X-band transmitter will consume about 11 W, which is quite high for CubeSat power standards. Nevertheless, the PULSAR team is working on developing its system by utilizing newer technologies including an X-band receiver and by improving the system's performance [19]. It is worth pointing out that [19] provides a detailed comparison between the performance of DSPs and FPGAs in handling SDR functions.

Maheshwarappa et al. [27], [28] proposed an SDR architecture based on an FPGA SoC and two A9 processors paired with an RF programmable transceiver SoC to support multi-CubeSat communications; reception of multiple signals using a single user equipment. The proposed system is not only meant for a portable ground station but also for an on-board CubeSat transceiver [27]. As it was previously mentioned, space SDR systems offer multiple features over conventional hardware design such as re-programmability during operation, flexibility to support multiple signals, and ability to support latest developments without hardware upgrade [28]. That is why [27] proposes a multi-core SDR architecture as shown in Fig. 3 to support multi-satellite communication. The baseband SoC contains an FPGA and two ARM dual-core Cortex A9 processors. The RF SoC used for evaluation was the Lime Micro Myriad RF containing the LMS6002D RF SoC, but the more recent version used AD-FMCOMMS3-EBZ containing the newer AD9361 RF SoC [27]. The samples received from the RF SoC are sent to the FPGA for processing. The FPGA is responsible for the more computationally intensive tasks including IQ correction, decimation, interpolation, and up/ down conversion. Modulation, packet handling, and other functions are performed on one of the A9 processors. Data and signal flow control for both the transmission and reception paths are performed on the other A9 processor. The FPGA and the processors are connected by high speed SoC Advanced eXtensible Interface (AXI) bus [27].

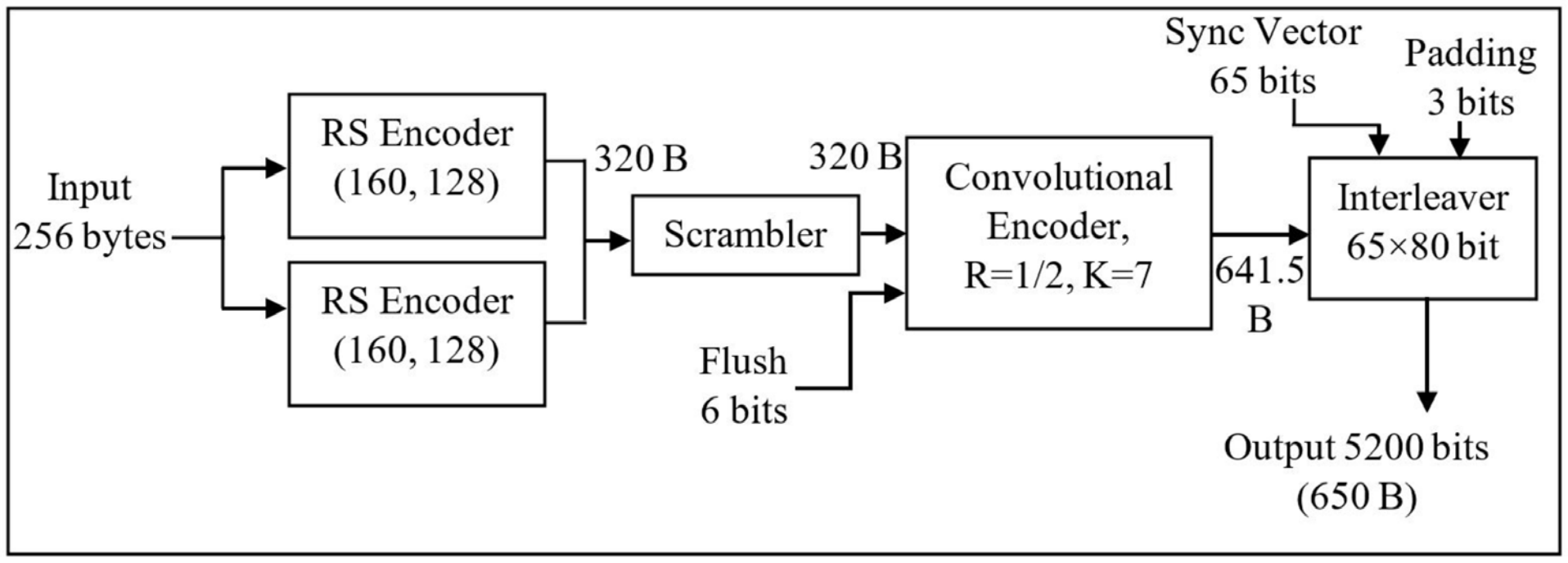

**Fig. 4.** FEC encoder based on Viterbi (rate ½) and two Reed-Solomon (160, 128) [27].

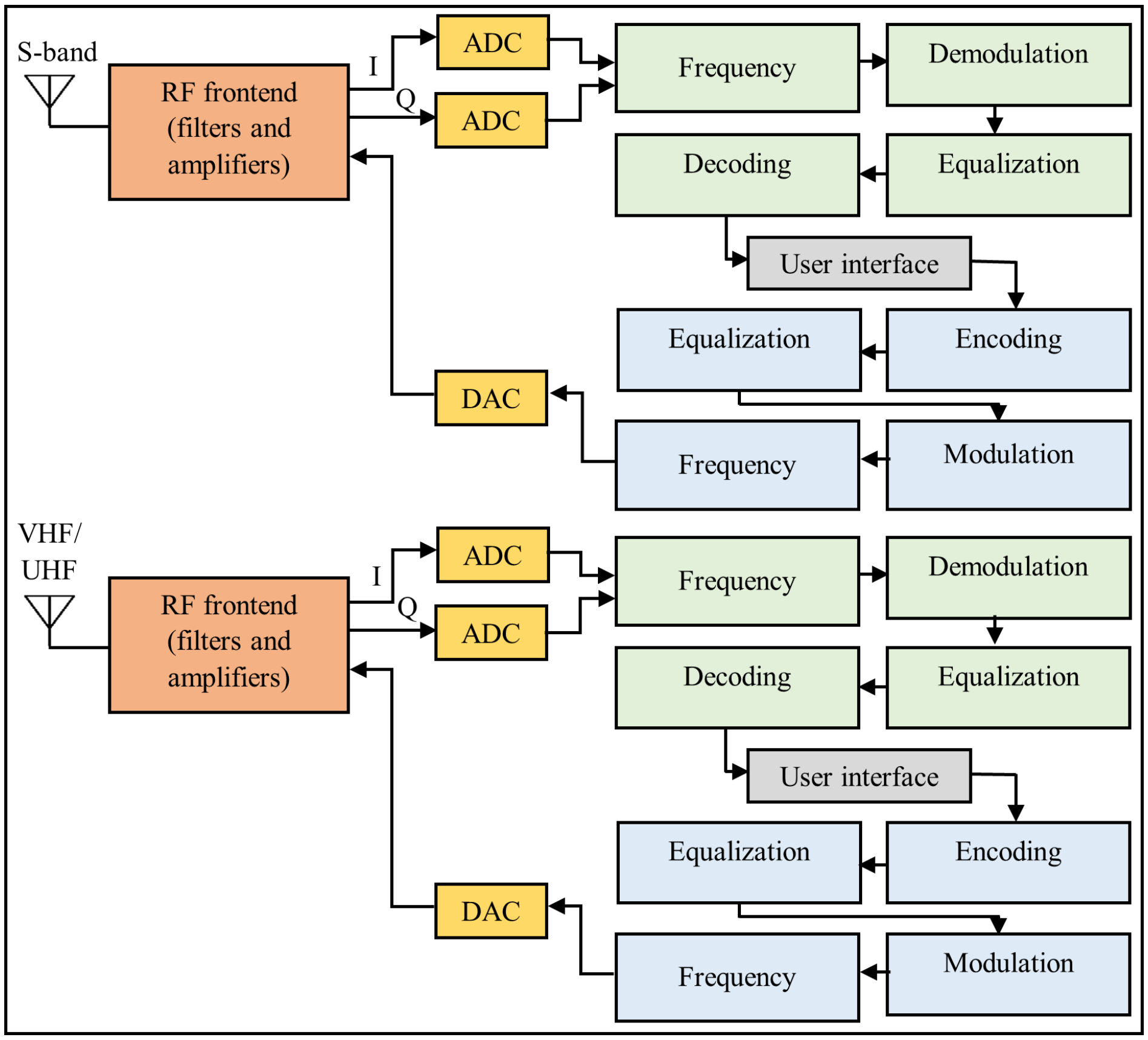

**Fig. 5.** Block diagram of the triple-band transceiver [28].

The modulation scheme used is BPSK [27], which is a commonly used modulation scheme among many CubeSats [29]. For the channel coding, an FEC scheme is adopted based on concatenated code using Viterbi (rate ½) and two Reed-Solomon (160, 128) blocks [27]. The working of this FEC encoder is illustrated in Fig. 4. Regarding the bandwidth utilization of this system, the VHF and UHF bands were used for uplink/ downlink while the S-band was used for inter-satellite link [28]. VHF and UHF were used for ground communications because there were more ground facilities using them thus increasing the communication window and because it was easier and cheaper to build VHF/ UHF ground stations [28]. While the S-band was chosen for inter-satellite communication to provide higher data rate [28]. The VHF/ UHF and S-band require different antennas and so separate RF front ends. As illustrated in Fig. 5, the SDR analog domain consists of bandpass filters for frequency selection, frequency conversion, and variable gain control. While the SDR digital domain is responsible for the other functions from encoding, equalization, modulation, to frequency/ amplitude/ phase offset correction [28]. The SDR system was first simulated using GNU-radio to have an understanding of the front end and back end blocks before hardware implementation. Two practical testbeds (SmartFusion2 and Zedboard) were then used to investigate the actual performance of the system under the expected space communications conditions by utilizing FUNcube-1 CubeSat and ESEO (European Student Earth Orbiter) microsatellite [28]. The initial testing on SmartFusion2 was to prove the SDR concept but it had many key SDR features missing, so the final testing was moved to Zedboard with Zipper and MyriadRF boards to overcome these problems [28]. The configuration of the second testbed was planned as in Fig. 3. The simulation and practical results are discussed in detail in [27] and [28]. The performance of the improved version of this SDR design, implemented on different platforms and employing different processors/ FPGA for various SDR tasks, is discussed in [30] and [31]. Overall, the system was able to handle up to 19.2 kbps at a memory requirement of 1.443 GB: consuming all memory on the Zedboard (256 KB), 560 KB extendable block RAM, and 1 GB external memory [31]. Higher data rates would thus require another board. The FPGA-based final implementation had a power consumption of 2.709 W [31]. This relatively high power consumption for the system's data rate is primarily due to the fact that the SDR is handling multi-satellite signals, working as a receiver for many satellites at the same time, and not only one signal as usually is the case. This in part accounts for its low data rate, relatively high power consumption, and very intensive computation complexity.

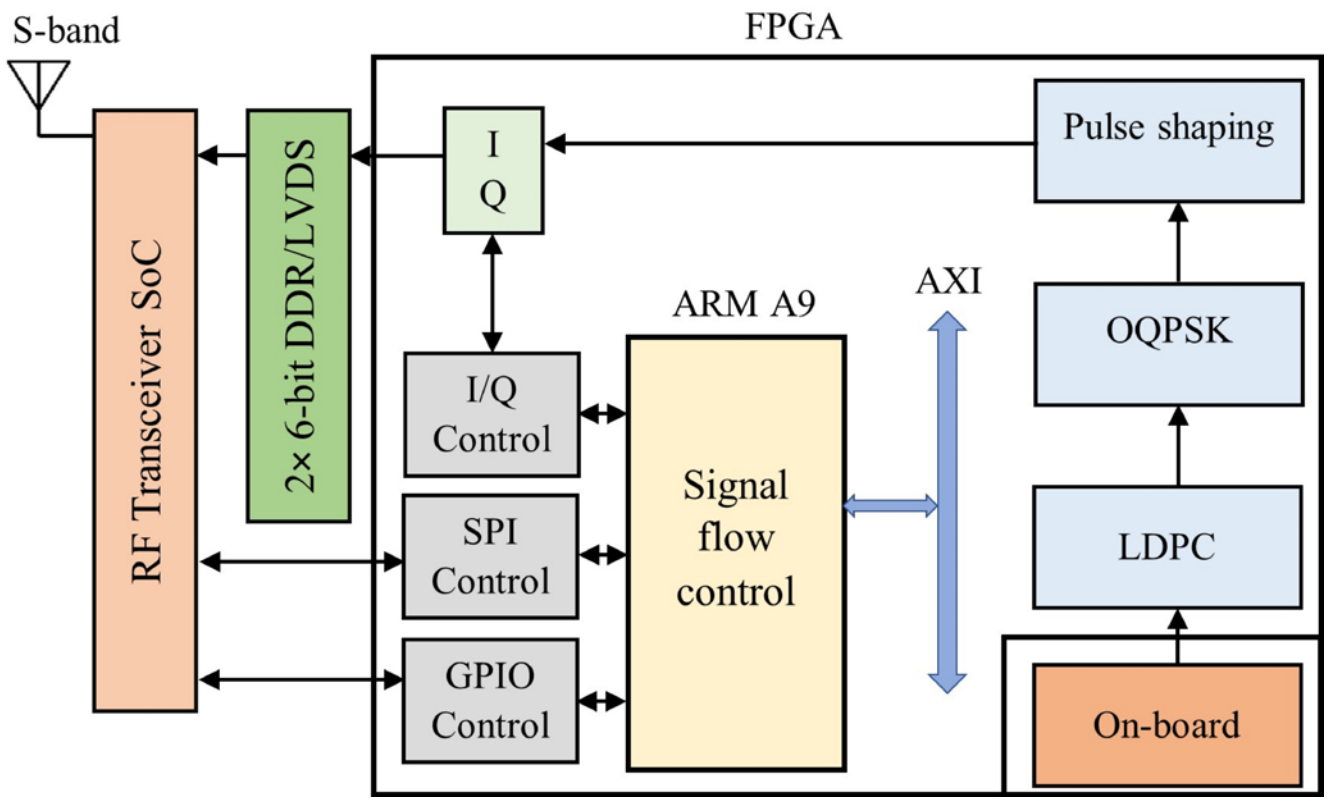

**Fig. 6.** Architecture of the SDR transmitter [32].

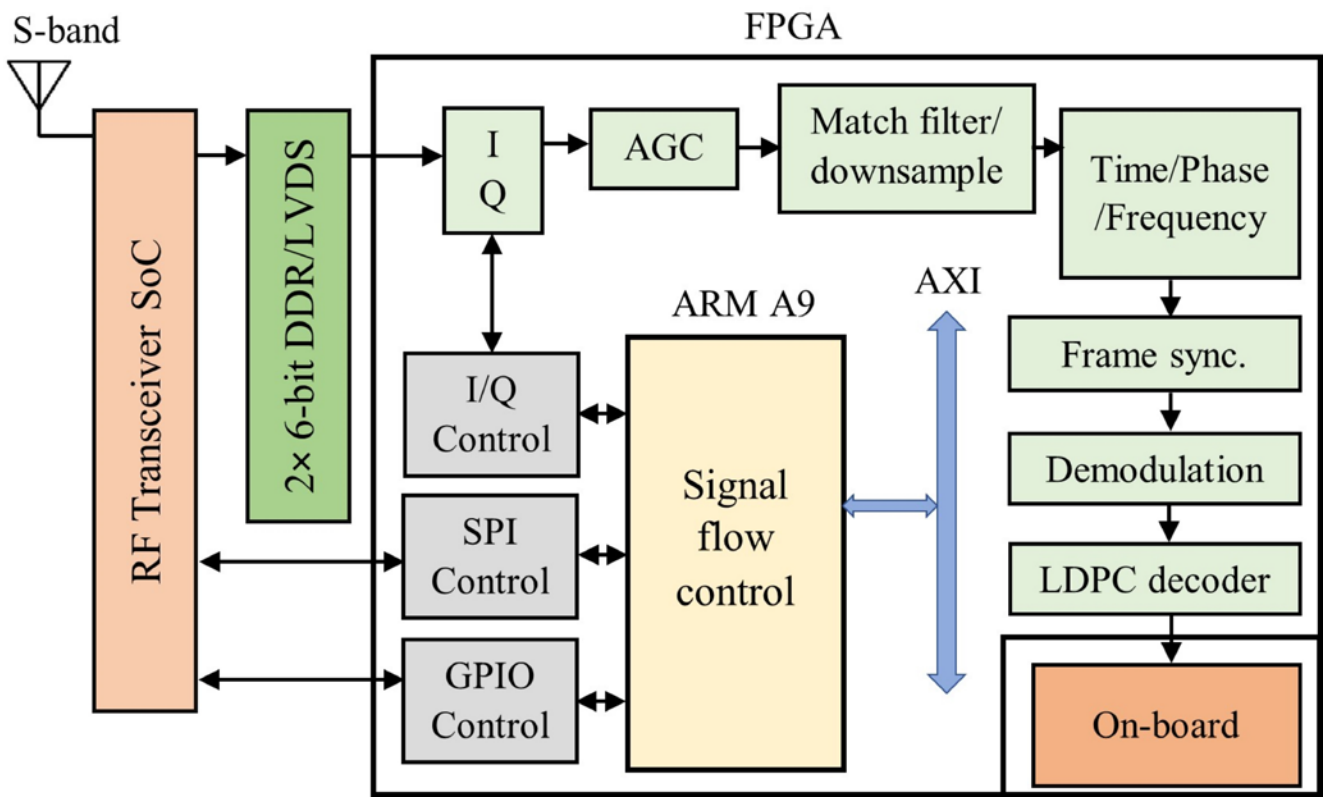

**Fig. 7.** Architecture of the SDR receiver [32].

Cai et al. [32] presented an FPGA-based SDR system for intersatellite communications (ISCs) suitable for small satellites. There were already previous ISC systems for small satellites such as [33]– [35] but they had considerable limitations. For example, [33] had a high power consumption to achieve a reasonably low bit error rate (BER) and [34] required around 50 iterations in the LDPC decoder alone to achieve acceptable error correction. Cai et al. use OQPSK modulation, and LDPC (255, 175) for their channel coding [32]. Due to the use of OQPSK modulation, an extra 0 bit is added to the code resulting in a code length of 256 bits. (n, k) LDPC is widely used because its capacity can approach the Shannon capacity limit and so for the same BER, the transmission power can be reduced [32]. A "dual-diagonal" matrix (a special form of the conventional parity check matrix) is used to obtain the codeword. The advantage of using this "dual-diagonal" format over the traditional one is that the number of computations needed in the encoding process is reduced from $(n^2 - k^2)/2$ to $(k+2)(n-k)$ [32]. Also, the number of "XOR" operations needed is reduced saving hardware resources. The pseudocodes for this encoding process as well as an improved version of it are illustrated in Algorithm 1 and 2 in [32]. The improved version requires even lower number of "XOR" operations resulting in even more efficient utilization of hardware resources. Although Algorithm 2 significantly reduces hardware utilization, Algorithm 1 provides higher throughput due to its parallel use of $(n-k)$ groups pf "XOR"; generates $(n-k)$ parity check bits simultaneously [32]. Fig. 6 illustrates the SDR architecture of the transmitter. An FPGA "Zynq 7020" was used for baseband processing to generate the different baseband signals and an RF transceiver "AD-FCOMMS3-EBZ" was used to transmit the RF waveform [32]. It can be seen that the FPGA is responsible for all baseband functions: LDPC encoding, OQPSK, and pulse shaping. In details, the data bits are generated by the PC then sent through an Ethernet interface to the FPGA which performs LDPC encoding on them and map the coded bits into OQPSK symbols to finally go through a square root raised cosine filter for pulse shaping [32]. On the RF front end, the pulse shaped signals are shifted in frequency by a 2.4 GHz carrier then transmitted. Fig. 7 shows the SDR architecture of the receiver, which shares the same structure with the transmitter. In this case, the RF signal is down-converted to a baseband signal by the transceiver and then sent to the FPGA to perform baseband processing: automatic gain control (AGC), matched filtering, demodulation, decoding, and other functions as illustrated in Fig. 7. Cai et al. present very detailed description of each of these blocks discussing the design, algorithms, and structure of each block. It was demonstrated that decoding with 50 iterations had almost the same performance as decoding with 20 iterations and only improves the performance by 0.1 dB at BER of $10^{-6}$ than decoding with 10 iterations [32]. Thus, with only 10 iterations the LDPC decoder can achieve very efficient error correction. Regarding the power consumption of this SDR design, the transmitter required a power consumption of 2.1 W while the receiver had a power consumption of 3.2 W [32]. Since the receiver is performing more computationally intensive tasks it had higher power consumption, especially that the transmitted power was quite low given that the transmitter and receiver were only separated by 2 m in the experimental test [32]. The data rate tested in the experimental setup was 122.88 kbps [32]. Although the data rate tested is low, supposedly it could reach up to 28 Mbps [32]. The main features of this design are its efficient LDPC algorithms, high code rate, and least $E_b/N_0$ at a BER of $10^{-6}$.

One of the most recent custom SDR systems for CubeSats is that of UOW (University of Wollongong) CubeSat [25]. The aim of the UOW 3U CubeSat project is to be able to transmit images from the satellite in the period of one pass over the ground station (60 seconds) while having an adaptive and on-flight reconfigurable communication system [25], [36]. For a maximum image size of 3 MB, the required data rate would be about 0.4 Mbps [36]. The SDR architecture is divided into digital and analog domain where the analog domain consists of an Analog Front End (AFE), transceiver, and RF front end. The SDR follows the architecture of a Zero-Intermediate-Frequency (ZIF) transceiver where there is no Intermediate Frequency (IF) stage in the communication system [36]. This results in a reduction of the hardware components needed and thus a highly integrated transceiver [25]. The transceiver selected is MAX2837 operating at a carrier frequency of 2.4 GHz [36]. Thus, the SDR operates in the S-band. Since this review is concerned with the baseband architecture, the details of the analog domain design will not be detailed and can be rather referred to in [25] and [36]. The digital baseband signal processing functions are performed using a Cyclone IV E FPGA from Altera [25]. 16-QAM modulation is used to provide high modulation efficiency [36]. This requires the use of two ADCs and two DACs; to operate on the in-phase component and out-of-phase component of the QAM signal simultaneously [36]. Fig. 8 illustrates the different baseband functions performed by the FPGA. The outgoing arrows, in Fig. 8, from the 3-wire SPI module is for the FPGA-based AFE control and transceiver control. The outgoing arrows from the time synchronization block go to a DAC block while the incoming arrows to the time recovery block come from an ADC block. Those baseband data converters make up the AFE stage. The MAX19713 is used for the AFE stage because it incorporates both a pair of 10-bit DAC and a pair of 10-bit ADC into a single device thus saving space without compromising the performance [36]. The device only consumes 91.8 mW at its full speed sample conversion rate of 45 MSps [36]. Using 16-QAM and three samples per each symbol with the MAX19713 device makes the maximum achievable data rate of the system 60 Mbps. Although this data rate is significantly greater than the minimum data rate needed for image transmission, it is not yet the actual data rate of the complete system since this depends on the performance of the other blocks of the SDR. Since this SDR operates in a half-duplex scheme, a limitation arising due to the chosen transceiver, only the transmit path or the receive path operates at a given time [36]. The output transmitted power of this system is 2 W, after using an external power amplifier at the transmission side in addition to the transceiver's preamplifier [25]. Since the received signal by the CubeSat from the ground station can be as low as -80 dBm [37], a Low Noise Amplifier (LNA) is used at the receiver's side to increase the SNR and minimize the added Noise Figure (NF) [25]. The chosen LNA has a gain of 13 dB and a very small NF of 0.85 dB at the carrier frequency [25]. Overall, the SDR has a power consumption of 0.4 W in the reception mode and 0.6 W in the transmission mode with an additional 2 W for the power amplifier so a total 2.6 W in the transmission mode [25], [36]. However, to optimize the power consumption of the system, the SDR has four operation states that will cycle through as it orbits the Earth. These states are shutdown, standby, receive, and transmit. The FPGA controls the operation state based on the position of the CubeSat in its orbit. When far away from the ground station, the SDR is in the shutdown state, it moves to the standby state when it approaches its communication window with the ground station, then it enters the receive state to receive an acknowledgment package from the station to initiate communication [25]. If the CubeSat has data to transmit, it enters the transmit state, otherwise, it remains in the standby state. This simple operation scheme considerably saves power when communication with the CubeSat is not possible, due to its orbital position, or not needed. This control is achieved using a finite state machine controller on the FPGA using the 3-wire SPI communication protocol [25]. It was found that the power consumption in the standby mode is 0.2 W while it is 0.03 W in the shutdown mode [25]. The fabricated PCB, having all the SDR components including the transceiver, AFE, amplifiers, and FPGA SoC, had dimensions of 9.5 cm × 8.8 cm, which meets the size constraint of a 1U CubeSat [25]. The SDR was successfully tested in the transmit mode and receive mode, but the testing had several limitations that are analyzed in the fourth section. It must also be noted that the SDR has no channel or source coding schemes implemented which will considerably affect the performance of the system under practical transmission and reception conditions.

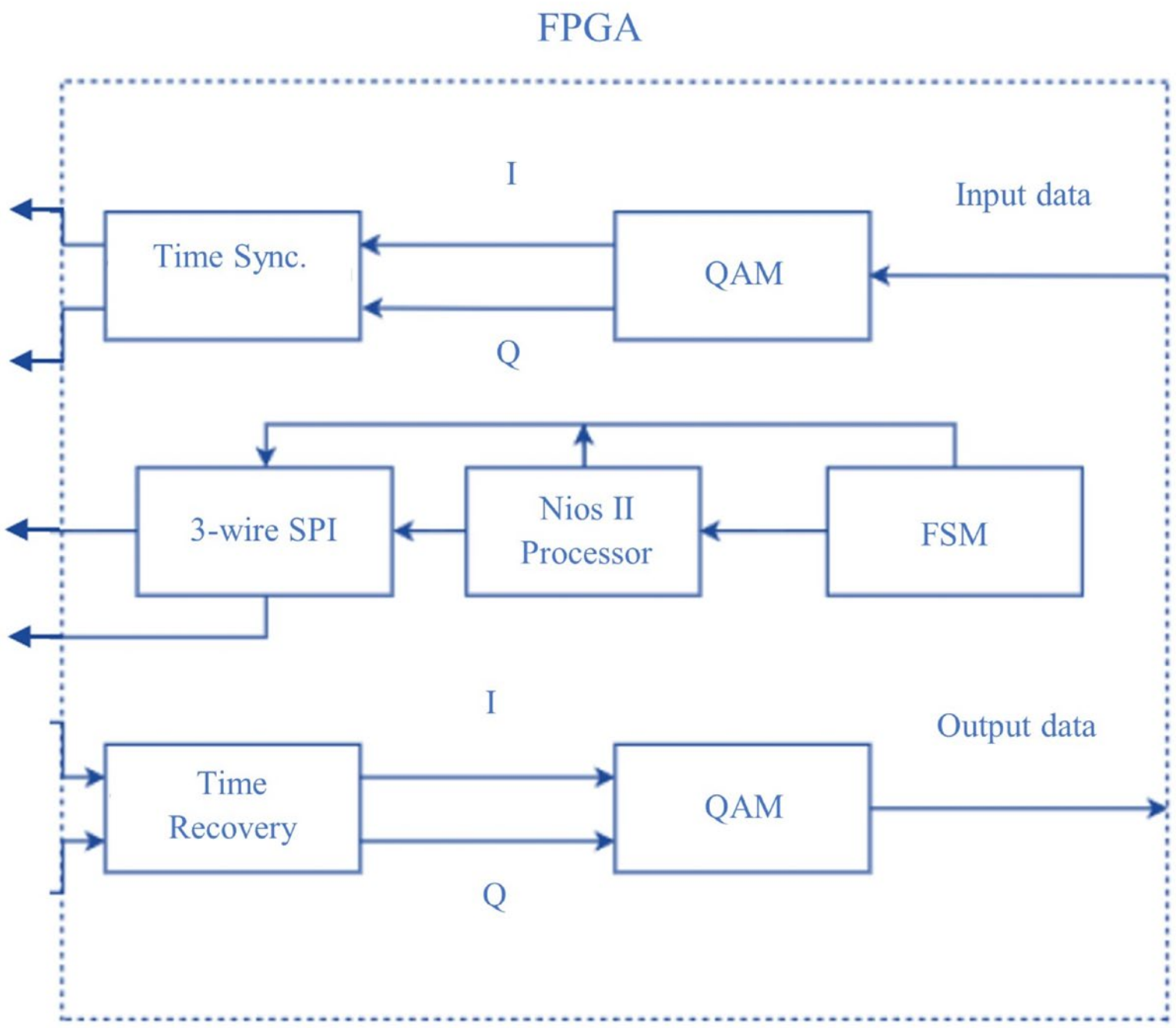

**Fig. 8.** Baseband architecture of UOW CubeSat SDR [25].

The final custom SDR system that will be reviewed is the AeroCube (Aerospace Corporation CubeSat) SDR system. There are several AeroCubes that have been launched, are currently in orbit, or are being developed. The design reviewed here is for AeroCube 915 MHz SDR communication system based on a Zynq processor [38-40]. It was found from previous AeroCube missions that as the CubeSat travels in its orbit, the SNR can change by more than 20 dB [38]. Consequently, adaptive coding and modulation (ACM) is employed in this system to track the link SNR value and optimize the data rate accordingly by changing the modulation and coding schemes used. The modulation schemes used in this design are BPSK and QPSK [38]. While root raised cosine is used for pulse shaping and Turbo (cdma2000) codec. Rate ¼, ½ are used for channel coding [38]. Using ACM over fixed modulation and coding schemes has the potential to more than double the average throughput of the SDR [38]. A Xilinx Zynq 7020 which combines an FPGA with two ARM Cortex-A9 processors was used to perform the baseband processing. For the RF front end, the LMS6002D field-programable RF IC was used in half duplex mode [38]. The frequency used by AeroCube is 915 MHz which lies in the UHF band. The developed SDR had a power consumption of 2.5 W in the transmission mode when transmitting a signal of 1 W, while the power consumption in the receive mode was 1.2 W [38]. To reduce the power consumption, the SDR is normally turned off and every 16 seconds it checks whether the ground station is attempting to establish a communication link [38]. In case a communication link is detected, the system will need 3 seconds to boot-up [38]. The data rate of this system is variable depending on the modulation and coding schemes used but based on the data and testing results presented in [38] it is around or less than 1 Mbps. The team planned on upgrading the system to use the Ka-band with an expected data rate of 10 Mbps at a transmission power consumption of 8.8 W and receive power consumption of 4.6 W [38]. However, it is not clear what is the status of that planned upgrade.

### *B. Commercial SDRs*

There are several CubeSat communication systems that are based on commercially available SDR platforms. Such platforms combine all the necessary hardware components (baseband and RF front end) to implement a completely functional SDR system that can be programmed and optimized to be used for a specific application including CubeSat communications. Examples of some widely used commercial SDRs are USRP E310, LimeSDR, SODA, and Iris. A comprehensive review of commercial SDR platforms can be found in [41] and [42]. While [41] focuses more on the SDRs' architectures and distinguishing features of each type of SDR platform, [42] is more of a practical guide for understanding and using these platforms and their supporting software packages. In many CubeSat projects, the commercial SDR platform is directly used in the system without any modifications or "design" from the CubeSat developer. The developer is solely concerned with deploying a functional CubeSat to achieve a given mission, rather than with designing a novel communication system that meets specific requirements, since this may not be the aim of the project. An example of such a system is that of HAVELSAT 2U CubeSat communication system. HAVELSAT design approach is to gather different subsystems from different commercial developers and integrate them to develop the HAVELSAT [43]. So, they directly used an SDR platform called GAMLINK developed by GAMANET company. The UHF band was used for downlink while the VHF band was used for uplink, and the SDR platform was immediately used in the CubeSat without intervention from HAVELSAT team [43]. Another example of this development approach is the USRP-based SDR system described in [44]. Such CubeSat designs are not part of this review since they are not concerned with the communication research-oriented design element that this paper aims to review.

$^3$CAT-2 is a 6U CubeSat that aims to demonstrate global navigation satellite system reflectometry (GNSS-R) by generating an observable called delay-Doppler maps [45], [46]. $^3$CAT-2 was launched in August 2016 [47] but is currently inactive since its lifetime has ended and because newer satellites ($^3$CAT-3 and $^3$CAT-4) are currently in development for launch [48-50]. Although the satellites have different scientific missions, their underlying baseband architecture, which was first developed for $^3$CAT-2, is almost identical [45], [48], [49]. The selected SDR platform is USRP B210 which comes without a case (board-only) making it easier to integrate with the rest of the CubeSat system [45]. USRP B210 features a Xilinx Spartan 6 FPGA, two transmitters and two receivers, fully coherent 2×2 MIMO, and ADC/ DAC resolution of 12 bits with a maximum sample rate of 61.44 MSps [51]. In order to enable two simultaneous receiving channels, which is required for the scientific payload, the SDR is used in dual-receiver mode. The USRP Hardware Driver (UHD) software is used to control the SDR [45]. The SDR here is not only used to communicate between the CubeSat and the base station but also to capture the data that is processed by the scientific payload of the satellite. Transmission (downlink) is carried over the S-band at 2100 MHz and the VHF band at 145.995 MHz while uplink is carried over the UHF band at 437.940 MHz [45]. Other frequencies are used for collecting the data in integration with the scientific payload structure which has its own antennas. Different modulation schemes and data rates are implemented in each of the used communication bands. The UHF band (uplink) employs AFSK modulation and has a data rate of 1.2 kbps [47]. Both the VHF and S-band (downlink) are modulated using BPSK with a data rate of 9.6 kbps over the VHF band and 115.2 kbps over the S-band [47]. The VHF and UHF links work together as a full-duplex system for telemetry and command upload while the S-band downlink is used for downloading data from the CubeSat without any uplink commands [47]. LDPC-Staircase is employed for error detection and correction. Data compression is also performed using a software called FAPEC which can perform lossless compression at a ratio of 1.5 and lossy compression up to a ratio of 40 [46]. However, the performed data compression is part of the scientific payload rather than the communication system. Based on the detailed link budget and power consumption data given in [46], the estimated power consumption of the CubeSat communication system is around 1.35 W. It is worth noting that $^3$CAT-3 communication system, which has different RF and antenna structure, has a much higher power consumption of 11 W for a slightly

improved data rate of 0.5 Mbps [48].

Alimenti et al. proposed a K/ Ka-band receiver that is suitable for deep space exploration missions, such as Moon exploration, based on off-the-shelf components [52]. The receiver is designed to be compatible with CubeSats or constellations of CubeSats that receive transmitted signals from the ground or from each other. The ground station is assumed to have a 10 m parabolic antenna and transmits a signal of 100 W power in the K-band at a frequency of 22.85 GHz [52]. It is further assumed that the transmitted signal is QPSK modulated and that the ground transmitter has a data rate of 100 Mbps, thus resulting in a signal of about 50 MHz bandwidth [52]. For the proposed receiver, a 50 cm diameter parabolic antenna is used [52]. It should be noted for some exploration mission scenarios, such as the one described in [52], more than one ground station with the above specifications may be needed for uninterrupted communication. Besides the RF and baseband frequencies processing, the proposed receiver has intermediate frequency (IF) processing functions at 3.7 GHz [52]. Therefore, the RF front end consists of an isolator, bandpass filter, two LNAs, and an image rejection filter. Then the signal is converted to the IF for further filtering and variable gain amplification. Finally, the IF signal is converted into a baseband signal and all baseband processing is performed on the commercial FPGA-based SDR platform. The SDR platform performs the demodulation of the QPSK signal (at IF) as well as the decoding and error correction based on the selected protocol which is DVB-S2 [52]. DVB-S2 is an abbreviation for digital video broadcasting – satellite – second generation [53]. It is a standard for satellite communications that defines a modulation and channel coding system suitable for a variety of satellite applications [53]. DVB-S2 supports a variety of modulation schemes including QPSK which is selected in this system. It employs LDPC coding concatenated with an outer BCH (Bose-Chaudhuri-Hocquenghem) code, with various possible rates, for its channel coding scheme [54]. Also, DVB-S2 includes a frame header that can be used for estimating the carrier offset due to Doppler shift [54]. So, overall, the developed SDR receiver employs QPSK demodulation and LDPC concatenated with BCH channel decoding schemes. The SDR is also responsible for generating a receiver strength indicator signal which is acquired by the receiver's CPU and used to implement an automatic gain control loop by adjusting the gain of the IF variable gain amplifier [52]. The SDR receiver is equipped with additional digital features which are autonomous Doppler shift compensation, autonomous handover between the different base stations, and autonomous determination of the uplink signal characteristics [52]. The method adopted for Doppler compensation allows the receiver to acquire the signal even if it has experienced extreme frequency shifts [52]. Regarding the handover, it should be clarified that the SDR receiver itself does not perform handover. In fact, it cannot control or initiate the handover [52]. However, it should be able to detect when the Earth base stations perform handovers. This can be realized in two ways: hard handover and soft handover. First, during the handover process, two base stations transmit the same signal simultaneously. In hard handover, the receiver separately detects the two signals and selects the most powerful one based on the link quality [52]. In soft handover, the receiver jointly detects the two signals and combines them after Doppler correction [52]. But soft handover is more complex as it requires special preamble or a cyclic prefix and additional processing at the receiver [52].

The proposed receiver is completely implemented using commercial off-the-shelf (COTS) components and evaluation boards. Its two main subsystems are the FPGA-based SDR modem (handles IF and baseband processing) and a Ka-band RF front end assembly [52]. The front end RF receiver is selected to work in the Ka-band at a frequency between 27.5 to 30 GHz to test the system's performance in the most critical scenario (highest frequency case) [52]. The RF QPSK modulated signal has a central frequency of 28.650 GHz, bandwidth of 50 MHz, and roll-off factor of 0.25 [52]. Although the SDR platform model is not specified, the FPGA used in the SDR is a Xilinx Kintex-7 device. The receiver is able to handle received signals with a data rate up to 80 Mbps [52]. However, since this design is only limited to an SDR receiver, the actual data rate of the complete communication system is determined by the ground stations rather than the CubeSat but, in any case, is limited to 80 Mbps. The paper presents a detailed analysis of the RF design, noise and amplifier calculations, and RF testing but it has very limited details about the baseband architecture of the system. This is mainly because the SDR modem handles all baseband processing according to the chosen protocol without much customization from the team. The total power consumption of the receiver is 8 W divided on the different subsystems as follows: 2.48 W consumed by the Ka-band RF front end, 2 W consumed by the IF SDR processing, 3 W consumed by the FPGA (baseband processing) and CPU, and 0.5 W dissipated by the power supply [52]. It should be kept in mind that this relatively high power demand is only for a CubeSat receiver; without a transmitter, which typically consumes more power than the receiver. Finally, the implemented receiver had a mass of 0.6 kg and dimensions of 9 cm × 9 cm × 4 cm [52], appropriate for CubeSat requirements.

Cadet is a relatively older CubeSat SDR that had a noticeably high data rate compared to other CubeSats at its time (2011) and even many recent CubeSats. The Cadet radio was developed in 2011 for the dynamic ionosphere 1.5 U CubeSat experiment, which consisted of two CubeSats communicating with two ground stations [55], [56]. Both CubeSats were successfully launched in October 2011 with Cadet radio providing a downlink data rate of 3 Mbps from each CubeSat to the two ground stations [55]. Cadet was completely implemented using power efficient COTS components. It works in a half-duplex mode using the UHF band, providing the first high speed communication system for a CubeSat [55]. The Cadet radio is fully controlled by the base station, it operates in the receive mode until it receives transmit command from the base station turning it to the transmission mode for a period of time and then it goes back to the receive mode. The downlink is carried over 460- 470 MHz while the uplink is carried over the 450 MHz frequency [55]. While the downlink data rate is 3 Mbps or more practically 2.6 Mbps with FEC employed, the uplink data rate is only 9.6 kbps since uplink is only used for the transmit command [55]. The uplink signal is FSK modulated while the downlink signal is OQPSK modulated, and some unspecified FEC scheme is used for the channel coding for both uplink and downlink [56]. The uplink signal is also encrypted using 256-bit AES (Advanced Encryption Standard) [55]. The uplink ground commands are sent by a TI CC1101-based transmitter and the downlink signal from the CubeSat is received using an Ettus USRP2 device [56]. Although the Cadet SDR processor type is not specified in the work, an FPGA is used for the scientific payload processing [56]. The Cadet SDR has a mass of around 0.33 kg and dimensions of 9.6 cm × 9.6 cm × 2 cm [55]. It has a power consumption of 141.6 mW for the receiver, 11298.0 mW for the transmitter (peak), and 30.0 mW consumed by the interface electronics, totaling a power consumption of 11.47 W [55].

### *C. Custom Hardware Designs*

After having reviewed SDR CubeSat systems, we move to systems that are mainly based on hardware components for implementing the different blocks of the communication system, which is the more conventional approach. As it was explained in the background, hardware implementation has some advantages over SDR implementation, and based on the specific application and method of design, it can have better performance in terms of power consumption, cost, and data rate. Therefore, it is important to investigate the status of hardware CubeSat communication systems and their performance in comparison with SDR systems to understand the unique features and limitations of each approach and the possible directions for moving forward. Although the use of COTS components for CubeSat communications is favorable in terms of cost, development time, and design complexity, reliability and efficient performance usually demand custom design hardware components. That is why many CubeSat projects follow the custom design approach for their CubeSats' communication systems. In this sub-section, we review the recent custom hardware CubeSat communication systems.

GeReLEO is a Ka-band satellite modem designed to provide connectivity to LEO satellites, via a data relay, with considerable constraints in mass, size, and power [57]. The Ka-band modem employs channel-efficient ACM with LDPC codes for error correction. Multi-frequency time-division multiple access (MF-TDMA) and multiplexing schemes are employed in the system [57]. The building blocks of the GeReLEO concept are LEO satellites equipped with a GeReLEO modem, a GeReLEO gateway, a GeReLEO network control center, and a data relay payload on-board a GEO satellite [57]. MF-TDMA is used for the data downlink and uplink whereas TDM is used for the low rate telecommand link [57]. One LEO satellite is serviced at a given timeslot over several frequency channels [57]. The duration of the timeslot is determined based on the visibility time of the satellite. The GeReLEO consists of a physical layer (PL) that generates the waveforms and a data link layer (DLL) that controls the time and frequency resources access. The GeReLEO transmission scheme is the key element for making the modem energy efficient, a limited complexity FPGA is used to control the processing of the system [57]. The modem supports 12 different modulation and coding schemes based on LDPC FEC and QPSK or 8-PSK modulation [57]. Furthermore, the modem has two codeword lengths: 9216 bit for the high-rate link and 2304 bit for the low-rate link [57]. The transmitted signal is a series of PL data frames where each sequence of PL frames starts with a synchronization frame, which consists of a string of alternating BPSK symbols and a code for frame synchronization. Each PL data frame in the sequence starts with a PL signaling header (7 bits) and contains either two QPSK modulated codewords or three 8-PSK modulated codewords [57]. The frame and codeword lengths of the high-rate link signal are four times the length of the low-rate link signal. The type of modulation and coding scheme used is specified using 4 bits in the PL signaling header of each frame [57].

The GeReLEO modem consists of the RF transceiver, IF blocks, and a Zynq SoC device. The digital signal processing functions as well as the DLL control and resource management are performed on the Zynq device. So, GeReLEO modem is more of a mixed software-hardware radio than a completely hardware implemented radio. The DAC, ADC, and filtering are performed in the IF stage at 70 MHz while the modulation and coding are performed on the Zynq SoC, specifically the FPGA [57]. The baseband architecture of the transmitter consists of three blocks as illustrated in Fig. 9. The first block is responsible for generating the header and data frames, the synchronization frame, and performing the LDPC encoding. The LDPC code rate is a variable parameter varying between 0.25 to 0.78 depending on the quality of the transmission link, and so it is set by the transmitting scheme [57]. This block also performs energy dispersal over the frames to have a uniform energy distribution [57]. The second block is responsible for the modulation: BPSK for the Sync frame and QPSK or 8-PSK for the data frame, depending on the link quality as determined by the transmission control software. It also combines the frame components into the PL frames according to the previously described structure. The third block handles signal shaping and generation. It performs digital pulse shaping with a roll-off factor of 0.35 then converts the signal to the IF and performs Doppler correction according to the calculated Doppler shift. Then, the DAC transforms the digital signal to an analogue signal at the IF and the signal is forwarded to the RF transceiver.

On the receiver side, the baseband architecture also consists of three main blocks as illustrated in Fig. 10. Firstly, the IF signal is digitalized, converted to the baseband frequency, demodulated, and filtered. Also, signal detection is performed concurrently to estimate the frequency, phase, and timing of the signal. A tracking algorithm running simultaneously to the next stage works to correct these estimates. Secondly, signal correction and interpolation are performed to correct the oversampled data in phase and frequency and downsample them. Thirdly, the symbols are divided into header and data symbols (demapped). The header symbols are processed to identify the start of the burst and the modulation/ coding scheme used. Furthermore, the quality of the transmission channel is evaluated using signal/ noise estimation performed on the header symbols [57]. Finally, the data symbols are LDPC decoded. The Ka-band carrier frequency used for transmission is 25.995 GHz and 23.040 GHz for reception, with a bandwidth of 36 MHz [57]. GeReLEO can achieve a maximum data rate of 1 Mbps in the low-rate link and a maximum data rate of 16 Mbps in the high-rate link [57]. The actual data rate in a given scenario depends on the selected code rate. Although the GeReLEO radio was not tested using the actual satellite configuration proposed for the system, an extensive testing procedure that simulates the actual configuration was developed to test the performance of the system, the testing setup and demonstration of the system are detailed in [57]. GeReLEO modem has a substantially high power consumption in 1U CubeSat power terms. Its digital board consumes 6 W, its DC-DC converter consumes another 6 W, and its RF front end consumes about 9 W totaling a power consumption of 21 W [57]. Moreover, its mass and dimensions are not suitable for 1U or even 2U CubeSats. The modem had a mass of 2.4 kg and dimensions of 20.3 cm × 20.0 cm × 7.7 cm [57]. So, it can only be suitable for 3U CubeSats or larger nanosatellites.

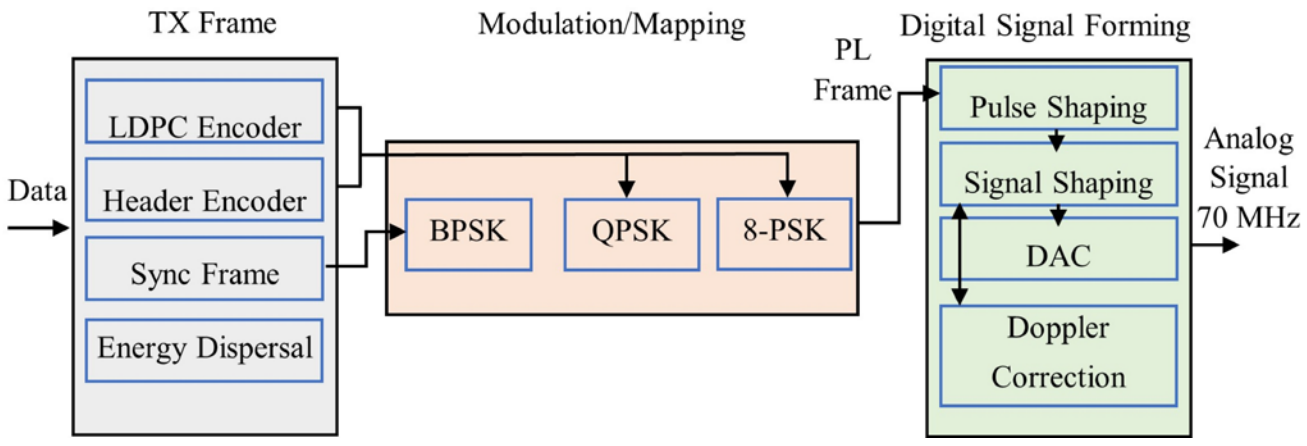

**Fig. 9.** Baseband architecture of the GeReLEO transmitter [57].

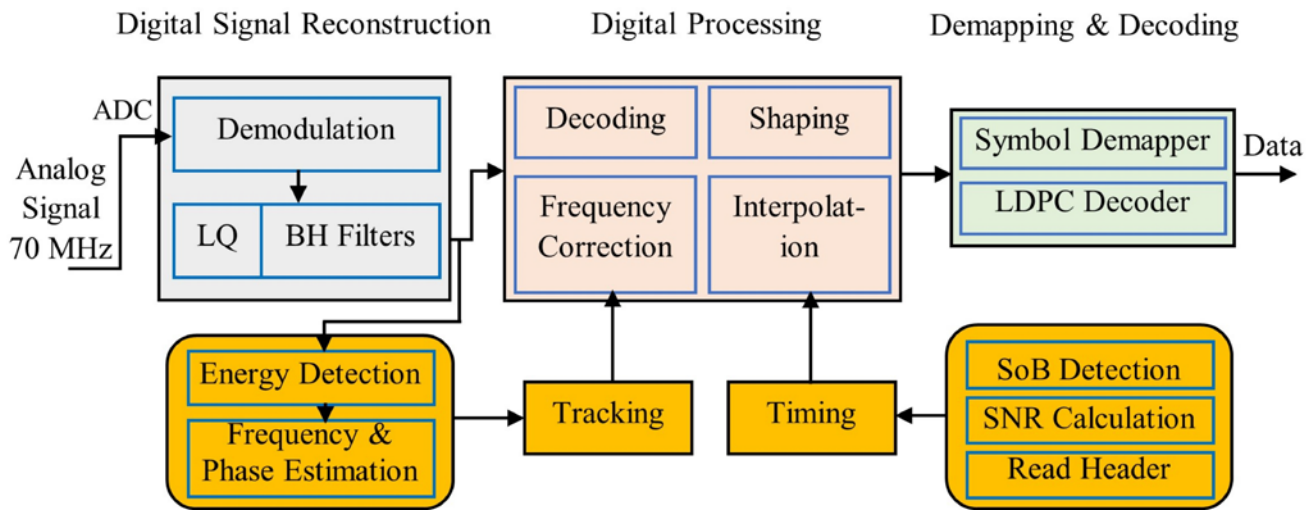

**Fig. 10.** Baseband architecture of the GeReLEO receiver [57].

The Joint Global Multi-Nation Birds Satellite (BIRDS) project is a CubeSat project that started with its first-generation constellation of five identical 1U CubeSats (BIRDS-1) in 2015 [58]. BIRDS-1 and 2 constellations were launched into orbit in the summer of 2017 [59] and 2018 [60], respectively. Moreover, BIRDS-3 and 4 were also successfully launched in 2019 [58] and 2021 [61], respectively. BIRDS-2, 3, and 4 are still active since their launch. The different BIRDS missions have different scientific and technology demonstration aims, and their underlaying systems, including the communication system, are improved versions from the designs of the previous generation. In other words, the 1U CubeSats of each BIRDS generation follow the heritage of the previous generation [58]. In particular, major developments in the communication system were realized in BIRDS-3 to establish a strong link between the satellites and the ground station overcoming the significantly lower performance of BIRDS-2 [58]. To comprehensively review the architecture of BIRDS communication system, we will start by reviewing BIRDS-2 system then discuss the improvements in design and performance introduced in BIRDS-3.

BIRDS-1 CubeSats employed two types of modulation with different data rates. AFSK modulation was used for sending control commands at a data rate of 1.2 kbps [59]. While GMSK was used for data uplink and downlink at a rate of 9.6 kbps [59]. The system used the UHF band for data link and the VHF band for the control line. The design was mainly concerned with implementing a customized UHF microstrip patch antenna, as described in [59], rather than designing all the blocks of the communication system. In BIRDS-2 communication system, the VHF and UHF bands are both used for data link using a commercial RF transceiver that handles APRS (Automatic Packet Reporting System) packets [62]. Any user that can handle APRS packets is able to receive transmitted signals from the CubeSats. The UHF band is used for data downlink whereas the VHF band is used for data uplink [62]. It must be noted that there are no intersatellite communications between the CubeSats of the constellation. The only communication links are with the 10 ground stations which are distributed over 10 different countries [63]. BIRDS-2 used the same modulation scheme (GMSK) and data rate (9.6 kbps) as BIRDS-1 for both uplink and downlink [63]. However, BIRDS-2 used deployable monopole antennas, whose design is described in detail in [64], instead of the patch antenna used in BIRDS-1 [63]. This transition to monopole antennas was made because the gain provided by the patch antennas was insufficient [62]. Also, monopole antennas are easier to deploy and have omnidirectional radiation pattern [64]. Similar to BIRDS-1 design, the main focus was on the design of the antenna and RF link while little attention was devoted to customizing the baseband architecture. Thus, the resulting communication system had simple design and low cost but required high power consumption and had a relatively large size. For example, the VHF monopole antenna had a length of 501 mm [64]. Consequently, BIRDS-2 faced many difficulties in its communication link that required major developments in BIRDS-3 communication system. Firstly, BIRDS-3 CubeSat uses a dipole antenna operating at the UHF band [58]. Moreover, several other improvements on the RF front end were developed such as improvement of the ground plane and EMI (electromagnetic interference) countermeasures [58]. On the ground station side, the RF transmitted power was increased from 14 W to 50 W [58]. Also, the antenna gain was increased by 4 dB by using circular polarized antenna instead of linear polarized antenna [58]. Overall, BIRDS-3 was able to achieve a link budget improvement of 17.7 dB compared to BIRDS-2. Although the improvements are mainly on the RF front end, there have also been improvements in the baseband design. On the baseband side, the uplink command size was decreased from 33 bytes to 14 bytes and the uplink speed was decreased from 9.6 kbps to 4.8 kbps since the amount of uplink data needed has been decreased and so slower rate can be used to increase the uplink success rate [58]. However, the downlink data rate has not been changed and GMSK modulation was used as before. Unfortunately, the detailed baseband architecture of BIRDS communication system is not described in any of the published works. However, based on the different references, it is clear that BIRDS use a very basic baseband architecture that has GMSK modulation without any source coding or channel coding schemes. BIRDS-3 CubeSat has a transmission power of 80.3 mW, after considering the antenna gain [58]. The precise power consumption of the communication system is not mentioned but based on the power supply description given in [62], the power consumption of the communication system is around 3 W. Overall, BIRDS-3 was able to successfully communicate with the ground station with a much higher performance than the previous BIRDS generations but without any improvements on the very low data rate of 9.6 kbps that was introduced since BIRDS-1. No major improvements in the communication system were reported for BIRDS-4 CubeSat.

Palo proposed a high-rate X-band transmitter suitable for CubeSat applications [65]. The proposed transmitter operates at a frequency of 8.380 GHz and has a transmitted signal power of 1 W from an omnidirectional antenna [65]. The X-band transmitter used has a power efficiency between 20-25%, which means that it consumes about 5 W to transmit a 1 W signal [65]. The transmitter has a basic baseband architecture that consists of OQPSK modulation and has two options for FEC, which are LDPC and convolutional encoding [65]. All other signal processing functions are performed in the RF frontend stage. An FPGA is used to generate random data to test the system. The data is first encoded using one of the FEC options then OQPSK modulated and then sent to the X-band transmitter to generate the RF signal and transmit it. However, before the baseband signal is sent to the RF stage, it goes through a sixth order analog lowpass filter that limits the signal's baseband bandwidth to ensure that the system meets the NASA emission spectrum requirements [65]. The proposed transmitter is able to achieve a data rate of 12.5 Mbps [65], however, the system has not yet been demonstrated in actual space environment and its testing was limited to the lab. The proposed transmitter is accompanied by a 200 kbps S-band receiver, but the design of the receiver is not part of the current work [65]. Given that the X-band transmitter has a power consumption of 5 W, and the baseband components have a power consumption between 1- 2 W, based on similar works, the total power consumption of the CubeSat transmitter is about 6.5 W. However, it must be noted that this figure does not take into account the receiver's power consumption, thus the power consumption of the complete communication system will be higher. Therefore, the main feature of this work is its potential to achieve a relatively high data rate for a CubeSat transmitter. However, the system is not thoroughly tested to confirm its error rate performance in an actual implementation scenario.

Phoenix is a 3U CubeSat that is developed to study the effects of Urban Heat Islands in several U.S. cities using infrared remote sensing [66]. The CubeSat is designed using some COTS components and several custom designed boards developed by the Phoenix team. Phoenix CubeSat was launched in November 2019 and was successfully deployed into its orbit from the International Space Station (ISS) in February 2020 [66]. The CubeSat was designed to meet a six-month mission baseline. Phoenix has three operation modes which are idle, science, and safe mode [66]. The idle mode accounts for around 90% of the orbital period of the satellite, during which the satellite is in low-power mode only transmitting and receiving telemetry commands such as the temperature and power status of the satellite [66]. In the science mode, the CubeSat performs its scientific payload operations such as thermal image collection and calibration [66]. In this mode as well the collected images are sent to the base station. Complete image downlink can require several passes over the ground station in order to transmit all the collected data. The safe mode is an emergency mode that is declared when a fault is detected in the CubeSat so that the CubeSat operation is stopped until the fault is fixed [66]. Both uplink and downlink are carried over the UHF band at 437.35 MHz [66]. A deployable UHF omnidirectional antenna developed by EnduroSat was used along with a GomSpace AX100 UHF transceiver [66]. The communication system employs GMSK modulation and uses the AX.25 data link layer protocol with HDLC (High-Level Data Link Control) encapsulation [66]. Additionally, cyclic redundancy check (CRC) is used for error detection and correction. Moreover, a custom packetization method was developed to help in recovering large files by the team. Before transmission, any file with a size greater than 256 bytes is broken into smaller files which are then transmitted as individual packets and reassembled into a single file in the ground station [66]. In case any packet is not received, it is re-requested from the CubeSat until the entire file is successfully reconstructed [66]. The communication system also utilizes an encryption scheme that is incorporated into all uplink command signals. The command signals are encrypted by a rotating cipher key that uses a simple substitution scheme [66]. Hence, during uplinks, the CubeSat will evaluate the received signal for a valid cipher passcode. If that passcode matches the expected passcode already known by the receiver, the uplink command will be accepted and executed accordingly [66]. The encryption process, and corresponding decryption, is only applied to the uplink signals from the base station. The downlink signals from the satellite which communicate the image data and all associated telemetry are not encrypted because they are not considered proprietary [66]. The communication system has a low data rate of about 10 kbps [66]. It is worth mentioning that the average power generated by the CubeSat's body-mounted solar panels during an orbit is around 6 W [66]. However, during Sun facing, the solar panels can generate up to 14 W [66]. Based on the simulation and testing of the CubeSat's power budget, the average generated power is sufficient to meet the average power consumption demand of the CubeSat. However, the exact power consumption of the communication system alone is not measured or specified. Although Phoenix CubeSat did not achieve its scientific mission due to an unexpected error following its deployment in orbit, it was able to establish a successful two-way communication link with the ground station. Thus, demonstrating the success of its communication system and allowing it to be used for calibrating ground stations to communicate with LEO CubeSats [66].

### *D. Commercial Hardware Designs*

In the final sub-section of our literature review, we review CubeSat communication systems that are primarily based on COTS hardware components. In this design approach, the CubeSat developer is not concerned with designing their own customized printed boards or ASICs but with making use of the most suitable commercial components that achieve the target of the CubeSat communication system in terms of the available energy resources, size, mass, cost, and anticipated data rate of the CubeSat based on its mission requirements. That is why commercial hardware designs can considerably vary in complexity and performance based on the target of the CubeSat developer and thanks to the huge variety of COTS components suitable for space applications. However, since the CubeSat developer is not much involved in the detailed design of the communication system, the information available on the architecture of the system can be very limited.

Planet, formerly Planet Labs, is an Earth imaging company that developed several CubeSat and small satellite constellations to image the whole Earth daily and identify environmental changes and trends. The company has different small satellite constellations, with different satellite designs, used for various Earth imaging missions. Planet operates the world's largest fleet of commercial remote sensing satellites [67]. The three main small satellite constellation projects of the company are the 3U Dove CubeSats (forming the Flock constellations), RapidEye, and SkySat [67]. In total, there are more than 24 Flocks, with more than 350 satellites, that have been launched by the company with different number of satellites in each Flock [67]. The average mass of the 3U Dove CubeSat is 4.7 kg [68]. The RapidEye constellation, which was retired in April 2020, consisted of 5 satellites each with a mass of around 156 kg and size of about 1 $m^3$ [69]. So, this constellation is not a CubeSat-based constellation and will not be part of the review. SkySat constellation, which is still active, is based on the CubeSat concept but with a scale up in mass and size. The constellation consists of 21 microsatellites each having a mass of approximately 100 kg and a length of 80 cm [70]. So, SkySat microsatellites will not be part of the review as well. We will only thus review the design and performance characteristics of the Dove CubeSat which forms the Flock constellations. The different Flock constellation missions and CubeSats are listed in [67].

Dove CubeSat is purely designed using COTS components that are integrated by the Planet team using their own circuit boards [71]. The components are chosen based on their performance, cost, size, and suitability for the Dove's mission which is Earth imaging [71]. Dove has a simple power conservative operation scheme; it starts by imaging the intended region and locally saving the images. Then, when the CubeSat is above the ground station, it automatically turns on its transmitter and downlink the images to the ground station. After the transmission process is completed, the transmitter is turned off again [72]. The operation of Dove is inclined towards using on-board fully automated systems for commissioning and operating the satellites rather than waiting for periodical commands from the ground station [72]. Manual commands are only used in case of anomalies or need of making system updates or corrections.

The communication system of Dove has considerably developed from the first to the current generation of Flock missions. The most sophisticated Dove communication system was designed for the 3U Build 14 Dove (B14) CubeSat, which was launched in November 2018 [73]. Dove B14 is the world's fastest X-band LEO satellite with an on-orbit maximum data rate of 1.674 Gbps [73], [74]. The satellite is able to download up to 85 GB of image data in one ground station pass [73]. Although the Ka-band has much more available bandwidth than the X-band, it has higher path loss through the atmosphere and suffers from limited availability of COTS components such as mixers and filters compared to the X-band [74]. Hence, Dove B14 employed the X-band for its high-speed radio, named HSD2 radio, since the design is completely based on COTS components. The radio consists of a CPU for control and processing, an SSD (solid state drive) for local data storage, six DVB-S2 modulation cores each running at a baud rate of 76.8 Msps for modulation and FEC, an FPGA for multiplexing the incoming data to the six modulator cores, two X-band transmitters, two antennas, and power amplifiers, filters, etc. [74], [75]. The majority of the modulation complexity is handled by the commercial DVB-S2 cores [75]. The radio operates in the 8025- 8400 MHz band and has dual circular polarization antennas that allow for doubling of bandwidth utilization [74]. That is why the radio has two independent transmitters and antennas. For each of the two types of circular polarization, three frequency channels are used simultaneously each with a bandwidth of 96 MHz and channel spacing of 100 MHz [74]. Hence, the use of six DVB-S2 cores as previously mentioned. Each individual channel has approximately a data rate of around 200 Mbps resulting in an average combined data rate of about 1.3 Gbps [74]. On the baseband side, adaptive modulation and channel coding is employed according to the DVB-S2 standard with modulation options of QPSK, 8-PSK, 16-APSK, and 32-APSK [74]. Regarding the channel coding, FEC based on LDPC concatenated with BCH is employed using code rates between 1/4 to 9/10 [74]. The DVB-S2 standard was selected due to the wide variety of options for the COTS components that can be used to realize this standard [75]. Even though there are specifically tuned standards for satellites such as those of the CCSDS (Consultive Committee for Space Data systems), those standards are not as practical as DVB-S2 due to the lack of commercial modems that can realize them [75]. The downlink is carried over the X-band with a maximum demonstrated data rate of 1674 Mbps [74]. On the other hand, the uplink is carried over the S-band at a data rate of 256 kbps [74]. On the ground side, there are five ground stations at five different locations with a total of 15 antennas [74]. Although the Dove B14 communication system has a very compact size of only 0.25 U, it has an exceptionally high power consumption of 50 W [74]. Such a high power consumption is expected given the intensive complexity of the system; six modulator cores, two transmitters, and two antennas all working simultaneously in addition to the FPGA and CPU. Thus, the very high data rate of the system comes at the expense of increased complexity and very high power consumption.

Another notable commercial hardware system is that of Corvus-BC CubeSat. Corvus-BC is a 6U CubeSat developed for multi-spectral Earth imaging [76]. The 1U communication system of Corvus-BC operates at a central frequency of 26.8 GHz with a bandwidth of 86.4 MHz in the Ka-band [76]. The Ka-band radio only works as a transmitter while telemetry and commands are sent over a separate UHF radio at low data rate [76]. The Ka-band radio employs the DVB-S2 link-layer protocol and is able to operate all of the 28 different modulation and coding schemes supported by the protocol [76]. The 28 different options are based on the valid combinations between one of the four modulation schemes, which are QPSK, 8-PSK, 16-APSK, and 32-APSK, and one of the 12 possible coding rates (1/4 - 9/10) of LDPC concatenated with BCH [54]. The maximum data rate achievable by this radio is 320.6 Mbps [76]. However, the average on-orbit data rate is about 185 Mbps [76]. The camera payload is connected via a Gigabit ethernet bus to the Ka-band transmitter. The images collected are losslessly compressed, encrypted, and streamed in packets over the Ka-band downlink [76]. Although the power consumption of the communication system is not specified, based on the typical power consumption of Ka-band CubeSat radios with similar data rates, it should be around 24 W [77]. One of the advantages of using the Ka-band is the smaller antenna size for both the transmitter and receiver. For instance, to achieve the same data rate over the X-band, the diameter of the ground station's dish has to be more than twice the diameter of the Ka-band dish which is about 2.8 m [76]. Larger dishes require considerably higher development and operation costs [76]. Moreover, the data rates achieved on the Ka-band, and the X-band as well, demonstrate these band's ability to achieve much higher data rates than the near term proposed CubeSat optical communication systems. In fact, the X-band and Ka-band

CubeSats achieved higher data rates compared to CubeSats using optical communication bands, and without the enormous technical obstacles that face these optical communication systems [76]. For instance, NASA's Optical Communications and Sensors Demonstration (OCSD) program was able to establish a CubeSat optical communication link at a maximum data rate of 100 Mbps using a 1.5U 2.3 kg CubeSat [78]. Although the achieved data rate is higher than most reviewed systems, the last two X-band and Ka-band designs achieved much higher data rates. Consequently, given the additional complexities and shortcomings of optical bands, such as the heavy dependence of optical connectivity on the cloud coverage, the X-band and Ka-band have much better potential in achieving very high data rate uninterrupted communication links than optical bands, so far.

Before closing this section, it is worth mentioning that there are some older (2000-2010) CubeSat communication systems that had some noticeable features such as the systems described in [79-84] which adopted different design approaches both hardware based such as [79] and software based such as [84], however, they are quite outdated for this review. For example, [84] was able to implement both AFSK and GMSK modulation schemes on a DSP connected to a VHF transceiver which had a transmission power consumption of 1 W for a data rate of 1.2 kbps for the AFSK modem and 4.8 kbps for the GMSK modem. The overall features of several such CubeSat communication systems that were developed between 2000 to 2011 are reviewed in [85]. It can be seen that the majority of those systems had data rates between 1- 10 kbps for a power consumption between 0.5- 1 W [85].

## IV. Performance Evaluation of the Reviewed Systems

### *A. Comparison Between the Reviewed Systems*

As it has been demonstrated through the previous section, there are various design approaches and methods for developing CubeSat communication systems. Although all the systems share the same general architecture of the digital communication system, they hugely vary in how they implement the different blocks of the communication system. While some designs go as far as developing their own improved algorithms and ASICs, others simply exploit existing standards and commercial components. Moreover, given that the different systems have been tested or demonstrated in different ways and under different conditions, it is neither straightforward nor accurate to fairly compare the systems based on a single metric. Accurate comparison is made even more challenging by the fact that some works have implicit or unstated assumptions in their design, testing, and consequently their results. Hence, the stated claims on the performance of the communication system may sometimes be misleading or inaccurate. Comparing the different systems in terms of their design approach, architecture, testing, and performance, however difficult, is both necessary and vital for moving forward in the development of high-performance more-efficient CubeSat communication systems. Consequently, this section is dedicated to providing a concise and accurate evaluation and comparison of the reviewed systems to better understand their features and limitations and identify directions for improvements, which will be discussed in the next section.

Table II provides a critical summary of the reviewed systems compared in terms of their design approach, mission objective, modulation and coding schemes, practical demonstrated data rate in Mbps, power consumption of the entire communication system in W, and the frequency bands used. Out of the 14 reviewed systems, 5 of them use the S-band, 4 use the VHF/ UHF bands, 3 use the Ka-band, and only 2 use the X-band, as their main transmission band. Although the review focused on low-power high-data rate systems, the power and data rate of the reviewed systems had considerable variation. Fig. 11 displays the data rate and corresponding power consumption of each of the reviewed systems illustrated by design category.

The y-axis, which represents the data rate in kbps, has a logarithmic scale to accommodate the orders of magnitude variation between the different data rates. Only 6 of the reviewed systems had a power consumption less than 5 W, 10 systems had a power consumption less than 10 W, while 4 systems had power consumption above 10 W. For the data rate, only 2 systems had a data rate above 100 Mbps, 4 systems had a data rate between 10- 100 Mbps, 2 systems had a data rate around 1 Mbps, and 6 systems had a data rate much less than 1 Mbps. It can be noted that 9 systems employ QPSK, either alone or with other types of modulation, and 8 systems employ LDPC for channel coding. However, only two systems employed some kind of encryption or source encoding scheme in their communication system. To conclude this comparison, it can be noted that more than half of the reviewed designs were based on SDR, specifically 8 systems, while 6 systems were mainly based on the more conventional hardware radio design approach.

### *B. Performance of Custom SDR Systems*

Although the PULSAR SDR system has the lowest power consumption among all reviewed systems and the highest data rate among all custom SDR systems, it did not go through extensive testing stages to demonstrate its reliability in the actual space environment. The FPGA-based design demonstrated very low power consumption of 1 W for a data rate of 10 Mbps but its performance in terms of BER has not been tested. However, given that the SDR utilizes various types of error correction schemes and NEN compatible packetization, it is expected that the system will have acceptable communication quality. On the other hand, Maheshwarapp SDR system has undergone extensive simulation, laboratory, and practical CubeSat communication testing stages. The system was tested as a ground transceiver using two different practical testbeds with both FUNcube-1 CubeSat and ESEO microsatellite and it successfully demonstrated reliable multi-CubeSat communication at a data rate of 19.2 kbps. Although the demonstrated data rate is very low with respect to the system's power consumption of 2.7 W, the SDR is not meant to provide high speed CubeSat downlink but to provide simultaneous multi-CubeSat signal reception capability. Consequently, the system has two separate communication chains working simultaneously over the S-band for inter-satellite link and over the VHF/ UHF band for uplink/ downlink as was illustrated in Fig. 5. This justifies the high power consumption to the low data rate of this system; it simultaneously receives and processes signals from multiple CubeSats. Therefore, although it may appear as if this SDR has low performance due to its data rate to power demand, it actually has quite high performance in terms of its energy-efficiency and communication reliability as a multi-CubeSat receiver. Cai SDR has the highest power consumption among all custom SDR systems. Moreover, its demonstration was only limited to in-lab testing and did not undergo actual space communication testing as opposed to Maheshwarapp SDR. Consequently, the BER and data rate of the system could considerably vary in actual CubeSat implementation from the values obtained during the in-lab testing. The most notable feature of this design is the improved algorithm of LDPC that was developed and implemented for the channel coding of this system resulting in very efficient utilization of the FPGA resources. Moreover, with this improved LDPC algorithm the achieved BER ($10^{-6}$) was much lower than other systems for the same number of LDPC iterations (20) and for the same SNR (4.9 dB). Thus, Cai SDR has superior BER performance compared to most other systems. Furthermore, besides the LDPC algorithm just described, Cai et al. also proposed another LDPC algorithm that uses more FPGA resources but has higher data rate than the reported data rate in Table II. Although it is claimed that the SDR can achieve a data rate of up to 28 Mbps, this data rate has not yet been demonstrated in actual testing. Although UOW SDR was tested in-lab, the testing lacked several important criteria. Firstly, the SDR was tested either as a transmitter or as a receiver at a given time. Therefore, there was no complete communication chain testing of the developed SDR where it would transmit to a certain receiver or receive a signal from another transmitter. As a result, the actual data rate that the system is capable of was not determined. Instead, only the maximum possible data rate limit (60 Mbps) allowed by the system's components is known. Although the actual data rate of this SDR was not determined, it is expected to be above 100 kbps based on the system's specifications and results from similar systems. Secondly, the testing was merely limited to verifying the functionality of the individual blocks, the overall functionality of the SDR, and the power consumption, rather than to test the practical performance of the communication system under realistic conditions. Thirdly, the testing was completely performed in-lab and no testing was performed using either existing CubeSats or UOW CubeSat itself. Furthermore, the SDR did not implement any channel coding schemes and so it is expected to have very high BER and correspondingly poor communication quality. Consequently, it is completely misleading to claim that UOW SDR has a data rate of 60 Mbps. Additionally, if any FEC scheme is going to be used in the system to make high-quality communication realizable, the power consumption will be significantly higher than the currently reported figure since coding/ decoding is usually computationally extensive. In contrast, AeroCube SDR, which had nearly the same power consumption as UOW SDR, was able to successfully implement FEC codes and even utilize ACM to change the modulation and channel coding schemes based on the channel conditions and in-orbit position. Moreover, AeroCube SDR demonstrated a data rate of around 1 Mbps based on in-orbit results. Although the BER performance of the system is not described, it is expected to have acceptable communication quality based on the overall successful results of the CubeSat.

TABLE II
COMPARISON BETWEEN THE REVIEWED CUBESAT COMMUNICATION SYSTEMS.

| System | Design Category | Mission Objective | Modulation Scheme | Coding Scheme | Data Rate (Mbps) | Power Consumption (W) | Bandwidth (GHz) |
|---|---|---|---|---|---|---|---|
| PULSAR [19] | Custom SDR (FPGA) | Technology demonstration | QPSK | LDPC, convolutional (rate ½), and Reed-Solomon (255/223) | 5- 10 | 0.5- 1 | S-band (2- 4) |
| Maheshwarapp et al. [27], [28] | Custom SDR (FPGA & ARM A9) | Demonstration of multi-CubeSat signal reception | BPSK | Viterbi (rate ½) and two Reed-Solomon (160, 128) blocks | 0.0192 | 2.709 | VHF (0.03- 0.30) UHF (0.30- 1.0) S-band (2- 4) |
| Cai et al. [32] | Custom SDR (FPGA) | Demonstration of CubeSat ISC | OQPSK | LDPC (improved algorithms) | 0.12288 | 5.3 | S-band (2.4) |
| UOW [25] | Custom SDR (FPGA) | Imaging | 16-QAM | None | Not tested (≥ 0.1, limited by 60) | 2.6 | S-band (2.4) |
| AeroCube SDR [38] | Custom SDR (FPGA & ARM A9) | Technology demonstration | BPSK, QPSK | Turbo (cdma2000) codec. Rate ¼, ½ | ~ 1 | 2.5 | UHF (0.915) |
| [3]CAT-2 [45-47] | Commercial SDR (USRP B210) | Global Navigation Satellite System Reflectometry (GNSS-R) | AFSK, BPSK | LDPC-Staircase | 0.1152 | 1.35 | VHF (0.146) UHF (0.438) S-band (2.1) |
| Alimenti et al. [52] | Commercial SDR (receiver only) | Deep Space Exploration | QPSK | LDPC concatenated with BCH | Not tested (≤ 80) | 8 | Ka-band (27.5- 30) |
| Cadet [55] | Commercial SDR | Dynamic Ionosphere Experiment | FSK, OQPSK | FEC. Encryption: 256-bit AES | 2.6 | 11.47 | UHF (0.450-0.470) |
| GeReLEO [57] | Custom Mixed (Hardware-Software) | Provide LEO inter-satellite links | QPSK, 8-PSK | LDPC (rate 0.25- 0.78) | 16 | 21 | Ka-band (25.995, 23.040) |
| BIRDS-3 [58] | Custom Hardware | Imaging & technology demonstration | GMSK | None | 0.01 | ~ 3 | UHF (0.437) |
| Palo [65] | Custom Hardware (transmitter only) | Technology demonstration | OQPSK | LDPC, convolutional | ≤ 12.5 | ~ 6.5 | X-band (8.380) |
| Phoenix [66] | Custom Hardware | Infrared remote sensing | GMSK | CRC. Encryption: Rotating/ Substitution Cipher. | 0.01 | < 6 | UHF (0.437) |
| Build 14 Dove (B14) [74] | Commercial Hardware | Earth imaging | QPSK, 8-PSK, 16-APSK, 32-APSK | LDPC concatenated with BCH (rate 1/4 - 9/10). | 1674 | 50 | S-band X-band (8.025-8.400) |
| Corvus-BC [76] | Commercial Hardware | Multi-spectral Earth imaging | QPSK, 8-PSK, 16-APSK, 32-APSK | LDPC concatenated with BCH (rate 1/4 - 9/10). | 320.6 | ~ 24 | Ka-band (26.8) |

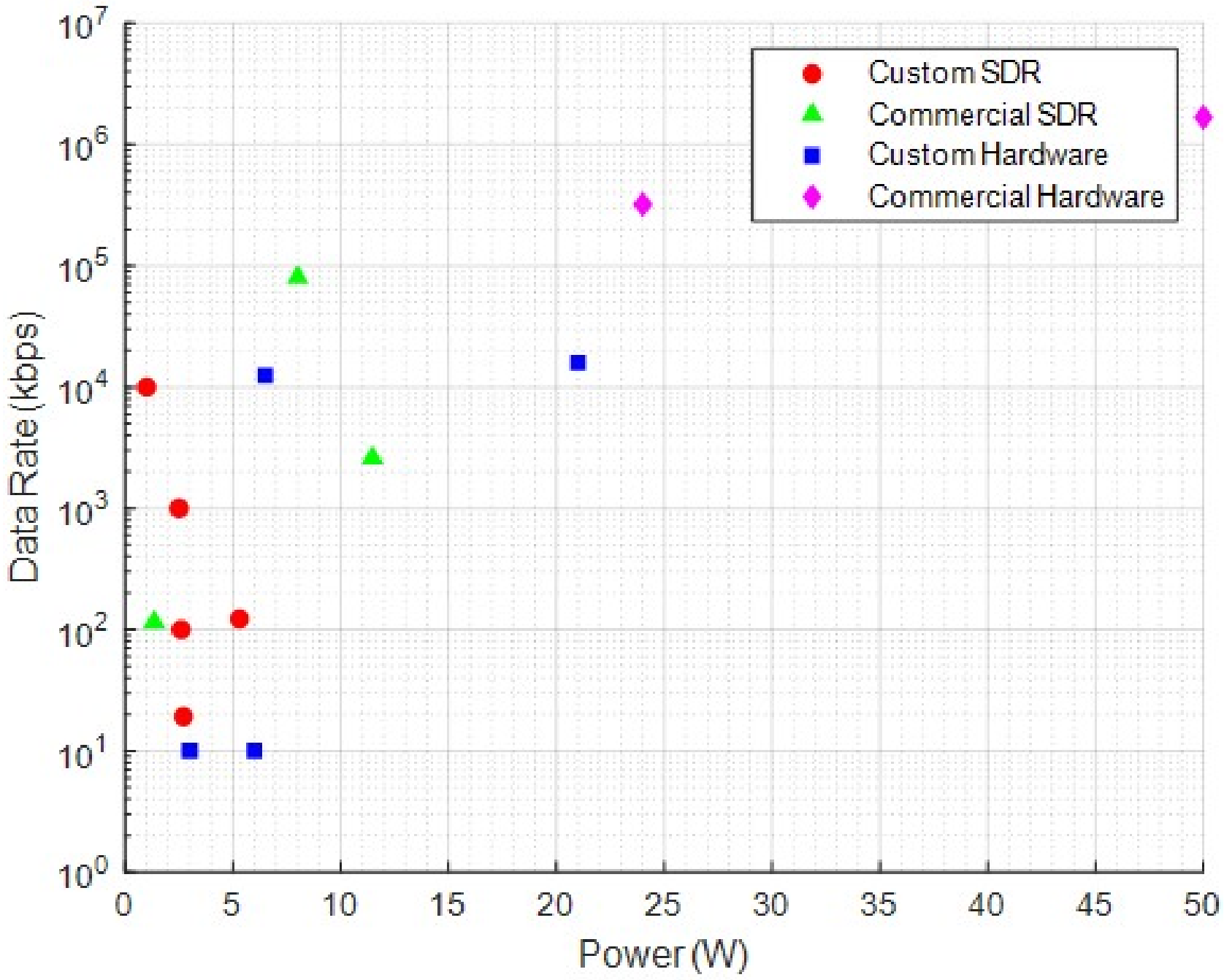

**Fig. 11.** Power consumption versus data rate of the reviewed systems by category.

### *C. Performance of Commercial SDR Systems*

Regarding commercial SDR systems, $^3$CAT-2 SDR had both the lowest data rate and power consumption among the reviewed systems. A distinguishing feature of $^3$CAT-2 is that its SDR platform (USRP B210) employs a 2×2 MIMO with two transmitters and two receivers. Therefore, the SDR is able to support two simultaneous transmitting and receiving channels. The SDR was successfully tested on board the $^3$CAT-2 CubeSat, providing a reliable 115.2 kbps downlink data rate over the S-band at an approximate 1.35 W power consumption. On the other hand, Alimenti SDR had a noticeably high power consumption for a receiver-only system. Although the Ka-band receiver is capable of handling up to 80 Mbps data rate, this has not been demonstrated in actual CubeSat deployment. Moreover, the data rate figure (100 Mbps) of the system provided in Table 7 of the paper [52] is quite misleading since it is the assumed data rate of the ground station not the data rate of the proposed receiver, which cannot handle data rates above 80 Mbps. Furthermore, the performance of the system has not been practically evaluated both in terms of the BER and data rate of the receiver in a complete CubeSat communication chain or under similar wireless channel conditions, most presented results are obtained from in-lab receiver-only experiments. Furthermore, due to the considerably high power consumption of the receiver-only system, around 8 W, it is very challenging to be used for typical CubeSat applications. This is because the CubeSat would require a transmitter that supports a similar data rate; for the receiver's high data rate capability to be utilized. However, transmitters usually require much higher power than receivers and thus the total power demand of the communication system will most likely be too high for usual CubeSat standards. Consequently, although the proposed receiver design is very attractive, it still probably needs to go through several testing and demonstration stages to compete with existing systems. Cadet SDR had the highest power consumption among all reviewed SDR systems. The SDR required a power of 11.47 W to achieve a data rate of 2.6 Mbps. A distinguishing feature of Cadet system is its use of encryption for uplink signals. All the uplink ground commands are encrypted using 256-bit AES and are decrypted by the CubeSat receiver. This ensures that the CubeSat does not respond to third-party commands. Although the exact BER performance of the SDR is not mentioned in the work, successful communication with the CubeSat was achieved demonstrating the reliability and performance of the system.

### *D. Performance of Custom Hardware Systems*

For the performance of custom hardware systems, it can be noticed that both BIRDS-3 and Phoenix systems had similar characteristics. Both systems employed the same frequency over the UHF band, both had a very low data rate of 10 kbps, and they both employed GMSK modulation. However, Phoenix had the major advantage of employing a channel coding scheme (CRC) whereas BIRDS-3 employed none. Moreover, Phoenix used a rotating cipher for encrypting all their uplink command signals. Due to these additional features of Phoenix system, it had higher power consumption than BIRDS-3 system. Nevertheless, BIRDS-3 had a more developed RF front end compared to Phoenix. For instance, while Phoenix used a simple omnidirectional antenna, BIRDS-3 used a circularly polarized dipole antenna with special EMI countermeasures. Although BIRDS-3 has a very basic baseband system with no channel or source coding schemes employed, it was able to accomplish its mission at its very low data rate. On the other hand, Phoenix failed to achieve its scientific mission due to an unexpected fault after its deployment, but it successfully demonstrated two-way communication according to the expected performance. However, both baseband designs are not competitive candidates for high data rate power efficient CubeSats. GeReLEO system had both the highest data rate and power consumption among the custom hardware systems. The design was not completely hardware-based but used SDR for several functions. A unique feature of GeReLEO system is the fact that it used multiple access techniques, which is obviously not common among the reviewed systems. It employed MF-TDMA for its data downlink/ uplink and TDM for its telecommand link. Moreover, the system passed through a rigorous testing stage that simulated the practical channel conditions for the proposed CubeSat configuration. The system demonstrated, in real hardware implementation, the ability to provide a high-quality reliable communication link with a BER of around $10^{-7}$ for an SNR of roughly 5 dB using QPSK with a code rate of 2/3. Given that GeReLEO system is the only custom hardware system that operates over the Ka-band and also the system with the highest demonstrated data rate (16 Mbps) among custom hardware systems, it is quite reasonable that it has the highest power consumption (21 W), greatest mass (2.4 kg), and largest dimensions (20.3 cm × 20.0 cm × 7.7 cm) among the reviewed custom hardware systems. Due to its power, mass, and size requirements, the GeReLEO modem is only compatible with 3U or larger CubeSats. Finally, Palo communication design consisted of a transmitter-only system. The system used a simple omnidirectional antenna with a transmitter power efficiency of roughly 25%. Although the proposed transmitter can achieve a maximum data rate of 12.5 Mbps, it has not yet been demonstrated on an actual CubeSat and its demonstration was limited to in-lab experiments, on the transmitter alone, with no details on how the system would perform in realistic space conditions. Furthermore, the proposed system, implementation, and testing results were all only concerned with the performance of the X-band transmitter, without enough consideration to the corresponding X-band receiver that will receive and decode the signal sent by the transmitter. In other words, testing was limited to verifying the functionality of the transmitter and not to demonstrating or evaluating the performance of the complete CubeSat communication system.  Hence, it is difficult to evaluate the proposed system in terms of actual data rate, reliability, and total power consumption. The BER performance of the system was also not stated in the work. Therefore, the system still probably needs to go through several design and testing stages before it can be implemented and deployed on a CubeSat mission.

### *E. Performance of Commercial Hardware Systems*

Commercial hardware systems had by far the highest data rates among all reviewed systems. Dove B14 demonstrated an in-orbit data rate of 1.674 Gbps making it the world's fastest X-band LEO satellite. This extremely high data rate comes at the expense of the very high, 50 W, power consumption of the radio, large mass of about 4.7 kg, and increased complexity of the communication system especially at the RF front end. It is worth mentioning that Dove's B14 power consumption is the highest among all reviewed systems and is slightly more than double the value of the second highest system which is that of Corvus-BC (24 W), also a commercial hardware system. However, Dove B14 has more than five times the data rate of Corvus-BC. Dove B14 was able to achieve such an exceptionally high data rate due to several reasons. Firstly, Dove B14 has two independent transmitters working at the same time. It has two antennas radiating simultaneously at two different polarizations (right-handed and left-handed circular polarization) and employs three frequency channels for each polarization. Thus, six simultaneous frequency channels are being used by the CubeSat for downlinking the data. Moreover, the communication system has several power amplification stages through the use of power amplifiers and from the high antenna gain of 12 dBi. Secondly, the system has six DVB-S2 modulation cores each operating at a baud rate of 76.8 Msps, simultaneously. Hence, each channel has its own high-speed modulation core. Thirdly, the system has a CPU dedicated to signal control and processing. Fourthly, the system has an FPGA dedicated to multiplexing the incoming data to the six modulators. Consequently, the communication system has superior processing power, high utilization of the available bandwidth resources, high gain, and employs different polarizations. Naturally, this results in very high power demand and large mass but effectively achieves the spectacularly high data rate of the system. Of course, the CubeSat has been successful in accomplishing its mission demonstrating the reliability of the communication system in terms of BER performance, in addition to the very high demonstrated data rate. Both Dove B14 and Corvus-BC communication systems employ the DVB-S2 standard. So, they have exactly the same modulation and channel coding options. However, while Dove B14 works over the S-band (uplink) and X-band (downlink), Corvus-BC works over the Ka-band at a much higher frequency. This has the key advantage of requiring smaller antenna size thus making the system more compact in size and mass. Similar to Dove B14, Corvus-BC also successfully accomplished its mission, demonstrating the reliability of its communication system at the stated data rate and power consumption.

## V. Discussion

To compare the general features of each of the four design approaches, Fig. 12 has been developed to provide a visual overview of how the different design approaches perform under eight different metrics. Under each metric, the performance of the approach is represented by the length of its arrow, where longer arrows indicate better performance. The metrics are the average consumed power over the achieved data rate of the communication system (average W per Mbps), the typical development cost, the required development time, the hardware and software complexity of the system, and the size and mass of the communication system. Longer arrows under these metrics indicate lower consumed power for a given data rate, lower cost, shorter development time, lower complexity, and smaller size/ mass, hence better performance. The other set of metrics are the average demonstrated data rate, the in-flight reconfigurability, and the capability of the system to be easily upgraded or scaled after deployment. For these metrics, longer arrows indicate higher data rate, better reconfigurability and scalability, and consequently better performance as well. The comparison is primarily based on the average performance of the reviewed systems under each design category. Hence, it reflects the general features of each of the design approaches for CubeSat communication systems.

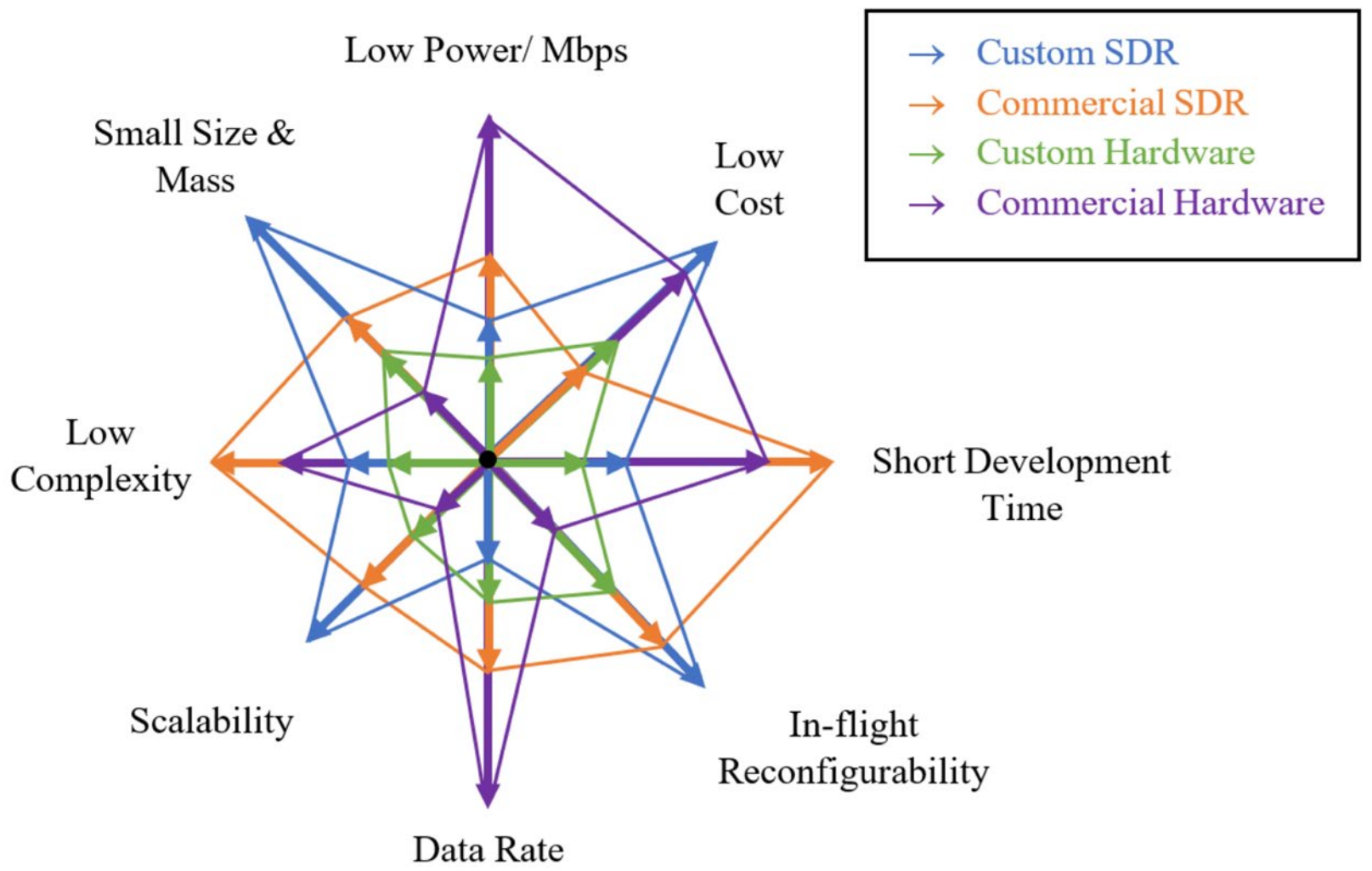

**Fig. 12.** Qualitative comparison between the features of each CubeSat communication system design approach.

It can be noticed that the custom SDR approach performs fairly well in most metrics. That is, it has the lowest cost of implementation, best in-flight reconfigurability, best scalability, smallest size, and at the same time it has moderate complexity and development time. Its primary limitation is that it has the lowest data rate and has a noticeably high power consumption with respect to the achieved data rate. However, this is in part due to the fact that custom SDR systems are still in development stages and not yet as widely used as commercial systems, which are usually preferred by large commercial companies and hence are more mature compared to existing custom SDR systems. With this fact in mind, custom SDR designs probably have high potential for realizing energy-efficient and high data rate communication systems that are suitable for CubeSats.

Commercial SDR systems also strongly compete with custom SDR systems, they have similar performance under several metrics such as reconfigurability, scalability, and size. However, they have slightly higher data rates and better power efficiency but at more expensive costs. Nevertheless, commercial SDR systems have the lowest complexity and shortest development time compared to all other systems. In contrast, custom hardware systems typically require the longest development time and have the highest complexity.

Commercial hardware systems succeeded in achieving the highest data rates ever achieved using CubeSats and small satellites in general, however, they have the largest sizes and masses among all other systems. Nonetheless, commercial hardware systems demonstrated the lowest power consumption required for achieving a given data rate. On the other hand, both commercial and custom hardware systems have very little capability to be reconfigured or upgraded after deployment. Consequently, they are not as flexible as SDR systems and if a fault were to occur in the system, most likely, it will be very difficult to fix, as actually happened with Phoenix CubeSat. On the other hand, SDR systems can be easily upgraded, reconfigured, fixed, and hence are less likely to fail the mission. This element of flexibility is a main motivation for many of the SDR-based systems such as the AeroCube SDR which is designed with the intention to be upgraded with more spectrally efficient waveforms and used as an in-orbit testing platform for new communication techniques before implementing them in newer versions of AeroCube satellites.

It is worth noting that while normally one would expect custom hardware systems to have smaller sizes and better power efficiency than SDR systems, this is not the case for the CubeSat systems considered in this review. This observation can be traced back to several reasons. Firstly, two of the reviewed custom hardware systems (BIRDS-3 and Phoenix) operated in the UHF band and had a considerably low data rate (10 kbps) for a relatively high power consumption, due to the specific system specifications (which were very similar for both systems) described in the literature review section and evaluated in the performance section. Consequently, this resulted in increasing the average power consumption per Mbps for custom hardware systems. Secondly, GeReLEO custom hardware system operated at a very high frequency in the Ka-band which largely explains why it had a very high power consumption (21 W) for its achieved data rate of 16 Mbps. Furthermore, GeReLEO communication system had a high mass of 2.4 kg and quite large dimensions. For example, this is in contrast with Alimenti commercial SDR system which also operated in the Ka-band but had a much lower power consumption of 8 W and a much smaller mass of 0.6 kg, however, Alimenti system was a receiver only. To summarize, the comparison given in Fig. 12 is mainly representative of the performance of the CubeSat communication systems reviewed in this paper and similar CubeSat systems, but it cannot be taken as a general comparison between the performance of software defined radios and hardware radios. In other words, the comparison must be interpreted in the context of this review.

Overall, Fig. 12 illustrates that commercial hardware systems had the highest power efficiency and demonstrated the highest data rates. On the other hand, SDR systems had the greatest flexibility and scalability. Moreover, systems based on COTS components (both SDR and hardware) had the lowest complexity and shortest time needed for development. Finally, it is concluded that each design approach has its own features and drawbacks. Consequently, the choice of which design approach to adopt largely depends on the specific goals and objectives of the CubeSat mission, a conclusion that is consistent with the CubeSat systems surveyed.

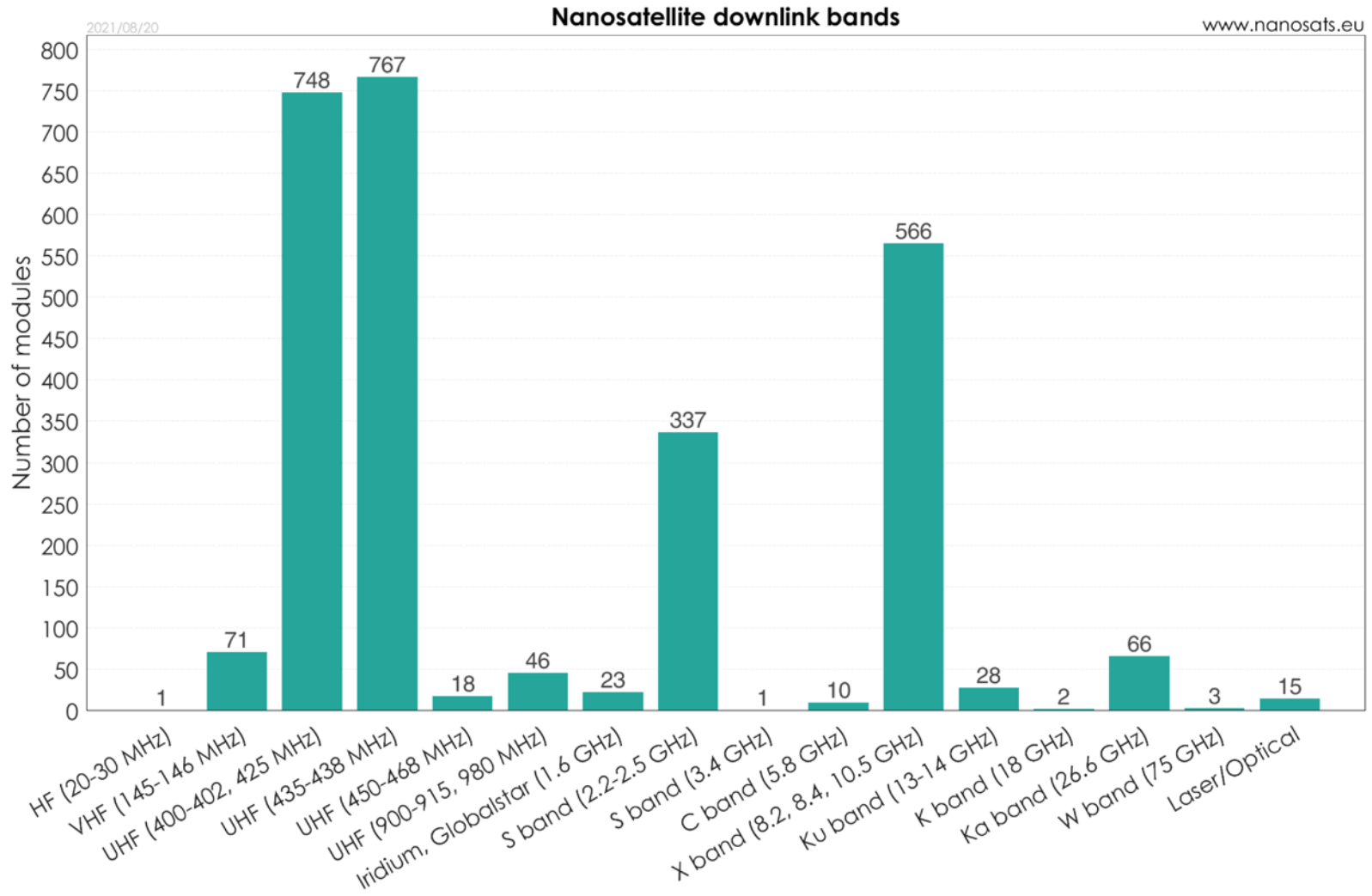

**Fig. 13.** Number of launched nanosatellites by downlink frequency band [8].

There are many directions for improving the capabilities of CubeSat communication systems to meet the increasing demands of higher data rates at lower costs, lower power consumption, smaller form-factors, and higher system flexibility. The improvements include using higher frequency bands, better modulation and coding schemes, improved baseband algorithms, use of MIMO (Multiple-Input Multiple-Output) and beamforming technologies, employment of multiple access techniques, use of advanced antennas, and use of efficient high-speed processors. Regarding the frequency bands, Fig. 13 illustrates the different bands used by nanosatellites launched since 1998. As it can be seen from the figure, the greatest number of nanosatellites communicated over the UHF band. As it has been previously mentioned, current and future trends include using higher frequency bands especially the Ka-band. Increasing the system's RF frequency not only increases the data rate, due to the more available bandwidth at higher bands, but also decreases the required antenna size and mass of the transceiver since the characteristic antenna size is directly proportional to the wavelength, or inversely proportional to the frequency. Due to this relatively recent trend towards using higher frequency bands, many nanosatellites are now operating over the S-band and X-band. Also, about 66 nanosatellites communicate over the Ka-band. It is expected that the number of nanosatellites operating at these high frequency bands will grow rapidly in the future. The transition towards higher bands is mainly driven by the demand for much higher data rates and the fact that low bands are highly congested. The bandwidth available at high frequency bands, such as the Ka-band, is many times larger compared to that available at lower bands and thus the achieved data rate can be many times higher [74]. However, this transition is hindered by the relative lack of energy-efficient COTS components operating at such high frequencies as well as the other challenges that face high-frequency communications such as increased power consumption, significantly higher path loss, and need for much faster processors to handle such high data rates. The Ka-band specifically has much higher attenuation through the atmosphere and higher rain and cloud fade [74]. Although the optical band is a very attractive option as it has practically unlimited available bandwidth, it still faces considerably severe challenges for CubeSat applications such as the very expensive costs, critical dependance of connectivity on cloud coverage, power efficiency of optical communication systems [74], as well as the other challenges discussed at the end of the commercial hardware designs sub-section.

Based on analyzing the data from a dataset [86] containing information on 757 CubeSats deployed in orbit as of 2018, Fig. 14 has been developed to give a statistical insight of the most common modulation schemes and antenna types used for the CubeSats that were launched between 2015 and 2018. The chart on the left displays the modulation schemes employed by 182 CubeSats launched during that period. It is evident that GMSK was the most widely used modulation scheme followed by BPSK, FSK, QPSK, and AFSK, which all had nearly the same percentage. Nevertheless, around 64% of the systems reviewed in this paper employed QPSK either alone or with other modulation schemes. Most CubeSat transceivers tend to use the aforementioned modulation schemes, especially GMSK and BPSK, due to the complexity of higher order modulation transceivers and the strict CubeSat limitations on energy consumption [87]. BPSK is the one of the most widely used modulation schemes for CubeSats because it is simple to implement and demands the least amount of power to support a given throughput [87]. GMSK is uniquely attractive for its efficient spectral characteristics, although it has worse BER performance compared to BPSK [87]. The chart on the right shows the most common antenna types used in CubeSat communication systems for 245 CubeSats launched in the same period. Dipole antennas and patch antennas were employed in almost 70% of the deployed CubeSats. As a matter of fact, the design of more efficient and high gain antennas is another critical method for improving CubeSat communications. For example, [88] describes the design and successful implementation of a high gain S-band slot antenna that significantly increased the gain from 2.52 dBi to 8.8 dBi. Improvements on the antenna design are not just limited to increasing the gain but can extend to decreasing the antenna size, increasing its power efficiency, and use of beam steering techniques if required [88]. Such improvements on the antenna side are a major future direction for improving the overall quality and power efficiency of CubeSat communication systems. Moreover, the use of multiple antennas, which is still not common in CubeSats, to exploit the features of MIMO and beamforming techniques is an important development to increase the data rate and serviceability, especially if we consider mobile ground stations that use CubeSats to provide communication services to users. Such development faces many challenges including power, size, and complexity constraints and the need for more powerful yet energy-efficient processors to execute the various MIMO/ beamforming algorithms. Consequently, employing beamforming requires considerable improvements on both the RF and baseband sides and that is why its use in CubeSats remains very limited. Nevertheless, its successful employment can significantly boost CubeSat communication capabilities and considerably increase the commercial applications of CubeSats.

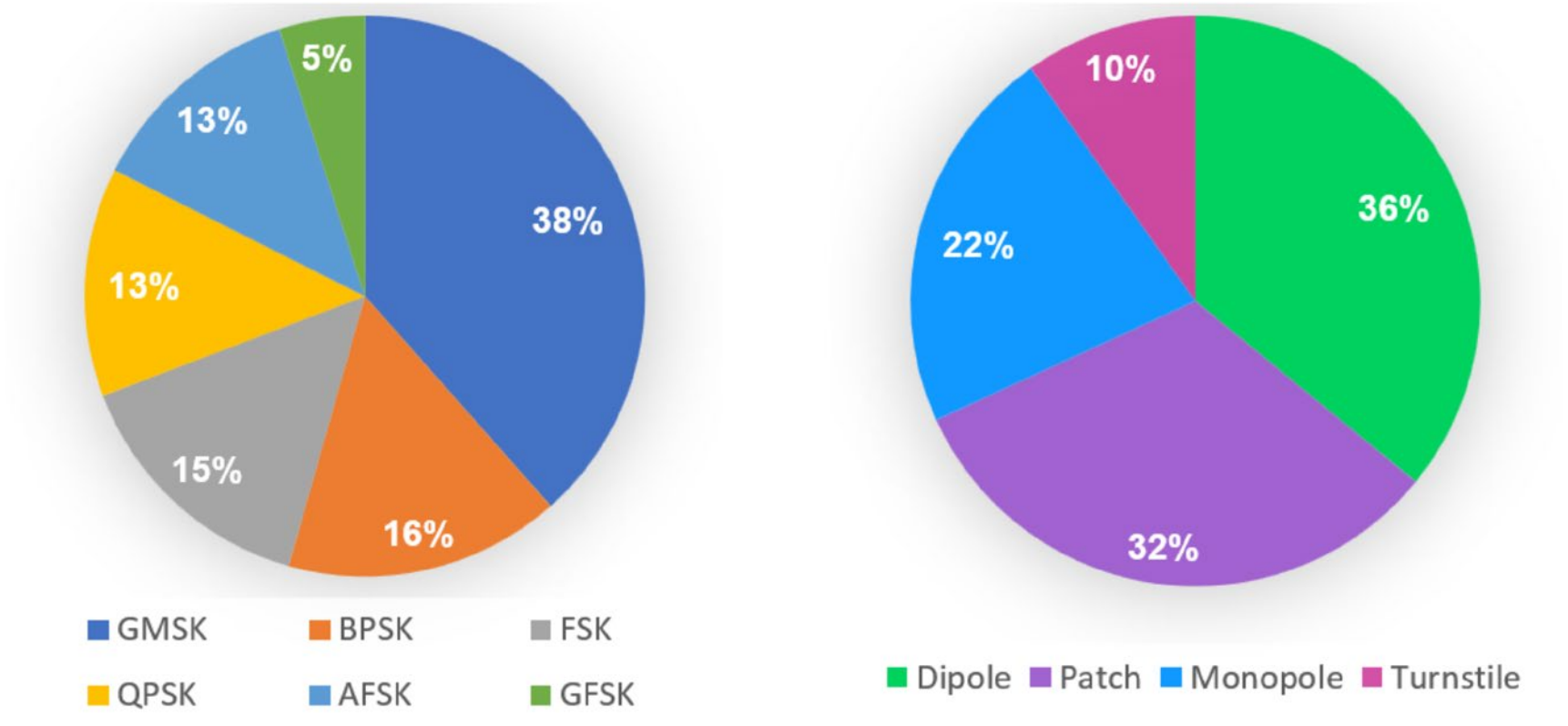

**Fig. 14.** Modulation schemes used by 182 CubeSats (left) and antenna type used by 245 CubeSats (right) launched between 2015 – 2018 [86].

Another aspect of future development is the use of multiple access techniques. Most current CubeSat systems do not make use of any multiple access techniques. Multiple access techniques such as FDMA (Frequency Division Multiple Access), TDMA (Time Division Multiple Access), CDMA (Code Division Multiple Access), and OFDM (Orthogonal Frequency Division Multiplexing) increase the efficiency of resource utilization and lead to higher data rates or larger number of serviceable users [89]. Implementing them on CubeSat systems will require higher power consumption but will correspondingly boost the performance of the system and the quality of the communication services that CubeSats can provide. Although multiple access techniques are still not widely used in CubeSats, some few existing CubeSats do employ them as was seen in the literature review and performance evaluation sections. Furthermore, the use of multiple access techniques for inter-satellite communication between small satellites was already long proposed, such as in [90] and [91]. The work in [92] presents a comparison between various multiple access techniques that can potentially be used to overcome some of the challenges that face inter-satellite communications. Therefore, even though there are several proposed architectures for employing multiple access techniques in CubeSat communication systems, the implementation of these techniques in current CubeSats is still quite limited.

On the baseband side, channel coding is an essential baseband block for all reliable CubeSat systems because it substantially reduces the required transmitter power by reducing the SNR requirement for achieving a given bit error rate. That is why almost all the reviewed systems employ some type of channel coding scheme. LDPC coding is the most used scheme as it provides a very high performance close to the Shannon capacity limit. Besides employing efficient error correction schemes, develop customized algorithms to implement the encoding and decoding processes is an important factor for designing more efficient systems. Improved channel coding algorithms can greatly reduce the required hardware resources needed to perform encoding/ decoding. This results in lower processing power utilization and hence lower power consumption. Moreover, channel coding algorithms can be customized to reduce the number of iterations required to achieve a specific BER at a given SNR or reduce the SNR requirement for a given BER. Furthermore, some algorithms can allow higher data rates than others based on the number of operations required and whether they are performed in series or in parallel. Consequently, developing improved and customized channel coding algorithms is one major way for improving the communication system in terms of efficiency, reliability, and data rate. Currently, only very few CubeSat systems go this far in developing and improving coding algorithms. This is reflected by the fact that out of all reviewed systems, only Cai's [32] system had improved algorithms for its LDPC coding. This was clearly reflected in its BER performance as was discussed in the previous section.

When it comes to source coding, it can be seen that most current systems do not employ any type of source coding scheme. While this can be understood by the motivation of minimizing the system's complexity and power consumption, efficient source coding can greatly enhance the performance of the system by reducing the size of the source data (data compression) and thus transmitting more information for a given data rate. This is especially useful when handling large size data such as images. In this case, source coding can greatly improve the system's performance. Image data compression was indeed implemented in some of the reviewed systems but to a small extent and not as part of the communication system design but rather as part of the scientific payload. Furthermore, the use of encryption was also very limited as it was only employed by two systems and only for their uplink commands. Needless to say, encryption would be necessary for CubeSats providing commercial, military, or any type of private communication service. Consequently, although the role of source coding may have been ignored in previous and many current CubeSat systems for the purpose of minimizing power and complexity, which is justifiable given the mostly research and technology demonstration applications, it certainly cannot be ignored for the uprising high-speed applications of CubeSats both from a performance perspective and a security perspective.

Of all the challenges that face CubeSat communication systems, the strict limitation on the allowed power consumption is the most severe challenge. The small size of the CubeSat restricts the size of its solar panels and batteries and thus limits the available power and energy resources. There are two approaches to overcome this limitation: maximize the harnessed solar power and minimize the systems' power consumption. There are several proposals for increasing the harnessed solar power such as the use of more efficient solar panels, use of deployable or foldable panels, and choice of optimum possible orbit that can maximize the generated solar power. When it comes to reducing the consumed power, of the communication system, this decrease should not come at the expense of lower performance. In contrast, the decrease in power should stem from using more efficient components that achieve the same performance at lower power consumption. One of the major aspects for realizing this is the use of FPGAs due to their unique energy-efficient features described in previous sections. Although FPGAs can provide high data processing performance at a reduced power consumption, FPGAs' power demand can still easily exceed CubeSats' power budget. For instance, the radiation tolerant Virtex4QV FPGA family's average power consumption can range from 1.25 W to 12.5 W, while a typical 1U CubeSat power budget ranges from 2 W to 8 W [93]. Even more, the maximum obtained power using body mounted solar panels on 3U CubeSats is typically less than 10 W [94]. Consequently, it is necessary to develop and employ energy reserve budgets and power-saving operation modes such as those described in the literature review section. Furthermore, [93] proposes two energy reserve budgeting scenarios for FPGA-based CubeSats based on the orbital pattern to determine the percentage of orbital time available for FPGA processes that require high power. Therefore, in order to minimize the power consumption of the communication system it is not enough to only use energy-efficient processors and components, the developer must also incorporate different operation modes and energy reserve budgets to optimize the power consumption of the communication system without compromising its performance. At the same time, there are several methods for increasing the generated solar power, most commonly by using deployable solar panels that are initially folded on the CubeSat sides and extend to their full size once the CubeSat is in its orbit. This technology is further improved by designing solar panels that can track the apparent motion of the sun. For example, [94] designed such a solar panel system for a 3U CubeSat consisting of two deployable systems made of three solar panels each for a total of six deployed solar panels that can track the sun's apparent motion. The system was able to deliver a maximum power of 50.4 W [94]. More generally, deployable solar panels produce between 160% to 400% more power than body mounted solar panels [95]. There are many deployable solar panel solutions for 3U CubeSats with total generated power ranging from 22 W to 56 W [96]– [98]. Therefore, using deployable solar panels greatly increases the amount of generated power which can be utilized to enable the employment of some of the techniques mentioned before such as beamforming and multiple access techniques and hence increase the data rate and boost the performance of the communication system. However, it must be kept in mind that these solar system solutions come at the expense of higher development costs, larger CubeSat mass, and increased complexity. Comparing the typical generated power, even with deployable solar panels, with the reported power consumption of the reviewed communication systems, keeping in mind that there are other systems on the CubeSat that also have high-power requirements, it is evident that there is a definite need for reducing the required power consumption per Mbps that the communication system can achieve.

## VI. Conclusion

This paper has presented a comprehensive and critical review of the design and architecture of CubeSat communication systems with a particular focus on baseband architectures. Four design approaches have been identified and the works falling under each category have been reviewed. The review focused on recent CubeSat systems that have relatively high data rate, above 100 kbps, and relatively low power consumption with respect to the achieved data rate. Commercial hardware communication systems showed the highest data rates reaching up to nearly 1.7 Gbps but also had the highest power consumption of up to 50 W. Custom SDR systems performed the best in terms of power demand and form-factor and had the unique advantage, along with commercial SDR systems, of allowing in-flight reconfigurability, upgradability, and scalability. Most reviewed systems employed QPSK modulation and used forward error correction for their channel coding. However, only two of the reviewed systems used encryption for the command uplink signals of their CubeSat systems. While most systems used already existing algorithms for their coding schemes, one of the FPGA-based custom SDR systems developed its own improved LDPC algorithms for better FPGA resource utilization, lower bit error rate for a given signal to noise ratio, and higher data rate. Only one of the reviewed systems used MIMO (2×2) technology in their CubeSat, which was a commercial SDR system. Moreover, only two systems used multiple access techniques, both were based on time division and/ or frequency division multiple access. The S-band and VHF/ UHF bands were the most commonly used bands among the reviewed systems, and generally among all launched CubeSats, while the X-band was the least common among the reviewed systems. However, there is an obvious trend towards moving to higher frequency bands for CubeSat communications as reflected from the number of reviewed systems that operates over the Ka-band and from Fig. 13 in the previous section which showed that around 680 nanosatellites are/ were operating over the X-band and higher frequency bands.

A detailed comparison and close evaluation of the performance, reliability, data rate, and power consumption of the reviewed systems were presented in the fourth section. It was found that some works overestimate the performance of their systems especially in terms of data rate. Furthermore, it was found that not all systems qualify as complete communication systems and some of them are still under development and require further testing stages before deployment. Moreover, it was seen that some works did not thoroughly investigate the bit error rate performance of their systems raising serious concerns about the reliability of their proposed designs.

CubeSat communication systems still face many challenges, namely the development of energy-efficient high-speed modems that satisfy CubeSats' cost, mass, size, and power requirements. The need for high-speed connectivity is becoming increasingly more essential with the rapid increase in the number of commercial CubeSat launches. Several directions for moving forward have been identified and discussed such as the use of improved coding algorithms, use of FPGAs, employment of multiple access techniques, employment of beamforming and MIMO techniques, use of advanced antennas, and transition to higher frequency bands. With such improvements and technology developments, the future of CubeSats seems very promising in playing a major role in the global wireless communication sector with applications ranging from Earth imaging and space exploration to military applications to commercial high-speed communications, smart cities, and IoT services and to many other uprising civil and commercial applications.

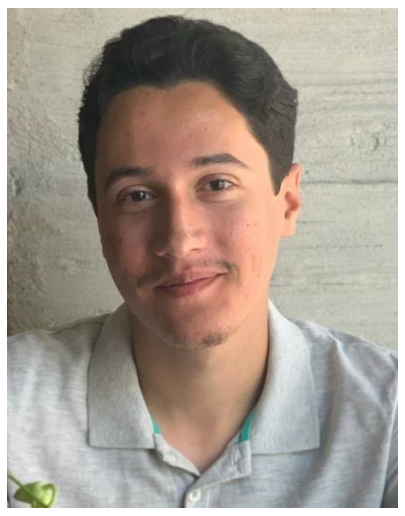

**Amr Zeedan** received his BS degree in electrical engineering from Qatar University, Doha, Qatar in 2021. He is currently pursuing his MSc degree in electrical engineering from Qatar University, Doha, Qatar.

He worked as an Undergraduate Research Assistant in the department of electrical engineering at Qatar University from 2020 to 2021. He is currently (starting from August 2021 till the present) working as a Graduate Research Assistant at Qatar University, Doha, Qatar. His research interests include IoT-based smart systems, signal processing, sustainable solar energy development, digital communication system design, small satellite communications, and quantum communication.

Mr. Zeedan has been awarded the Amiri Undergraduate Academic Excellence Scholarship during his undergraduate study. He has also been awarded the Graduate Sponsorship Research Award (GSRA) from Qatar National Research Fund (QNRF) covering the two-year period of his current graduate program.

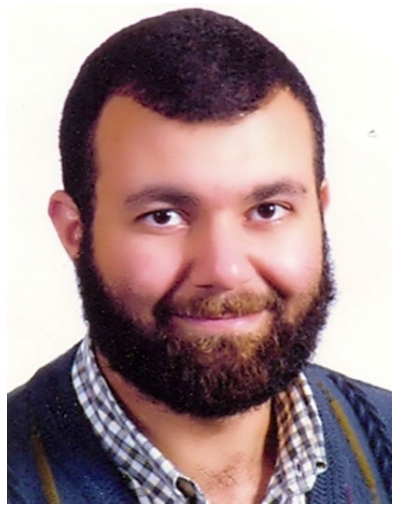

**Tamer Khattab**, (Senior Member, IEEE) received the B.Sc. and M.Sc. degrees in electronics and communications engineering from Cairo University, Giza, Egypt, and the Ph.D. degree in electrical and computer engineering from The University of British Columbia (UBC), Vancouver, BC, Canada, in 2007. From 1994 to 1999, he was with IBM WTC, Giza, as a Development Team Member, and then as a Development Team Lead. From 2000 to 2003, he was with Nokia (formerly Alcatel Canada Inc.), Burnaby, BC, Canada, as a Senior Member of the technical staff. From 2006 to 2007, he was a Postdoctoral Fellow with The University of British Columbia, where he was involved in prototyping advanced Gbits/s wireless LAN baseband transceivers. He joined Qatar University (QU) in 2007, where he is currently a Professor of electrical engineering. He is also a Senior Member of the technical staff with Qatar Mobility Innovation Center (QMIC), a research and development center owned by QU and funded by Qatar Science and Technology Park (QSTP). In addition to more than 150 high-profile academic journal and conference publications, he has several published and pending patents. His research interests include physical layer security techniques, information-theoretic aspects of communication systems, radar and RF sensing techniques, and optical communication.